**FRONT MATTER**

# Title

Long title:
Room-temperature superconductivity in heavy rare earth metal substituted sodalite-like clathrate hexahydrides under moderate pressure
Short title: Superconductivity of Yb/Lu substituted clathrate hexahydrides

Please provide both a long and a short title for your paper.
Full titles can be a maximum of 200 characters.
Short titles can be a maximum of 60 characters.

# Authors

Mingyang Du[1], Hao Song[1], Zihan Zhang[1], Defang Duan[1]*, Tian Cui[1,2]*

# Affiliations

1 College of Physics, Jilin University, Changchun 130012, People's Republic of China
2 Institute of High-Pressure Physics, School of Physical Science and Technology, Ningbo University, Ningbo, 315211, People's Republic of China
*E-mail: duandf@jlu.edu.cn, cuitian@nbu.edu.cn

# Abstract

Room temperature superconductivity is a dream that mankind has been chasing for a century. In recent years, the synthesis of $H_3S$, $LaH_{10}$ and C-S-H system has gradually made this dream a reality. But the extreme pressures required for the metallization of hydrogen-based superconductors limit their applications. In this work, we design a series of high temperature superconductors that can be stable at moderate pressures by incorporating heavy rare earth elements Yb/Lu into sodalite-like clathrate hydrides. In particular, the critical temperatures of $Y_3LuH_{24}$, $YLuH_{12}$ and $YLu_3H_{24}$ are 283 K, 275 K and 288 K, respectively, which are close to or have reached room temperature, and the required pressure for stabilization of these hydrides is about 120 GPa which is significantly lower than that of reported room temperature superconductors. Our work provides an effective method for the rational design of low-pressure stabilized hydrogen-based superconductors with high-$T_c$ and will stimulate further experimental exploration.

## MAIN TEXT

### 1. Introduction

Since H. K. Onnes observed superconductivity in Hg in 1911, room-temperature superconductivity has been a long-sought dream and a field of intensive research. E. Winger and H. B. Huntington theoretically predicted in 1935 that solid metal hydrogen can be obtained under high pressure conditions[1]. According to the BCS theory, the superconducting transition temperature of material is proportional to its Debye temperature[2]. This suggests that hydrogen, the lightest element in nature, would be an ideal room-temperature superconductor after metallization[3]. However, experimental studies have shown that hydrogen requires extremely high pressures to metallize[4, 5]. As a result, the search for room-temperature superconductor has gradually turned to another, more feasible route — hydrogen-rich compounds.

By incorporating other elements to create a "chemical pre-compression" effect on hydrogen, hydrogen-rich compounds can be metallized at much lower pressures than pure hydrogen[6]. Guided by this principle, many excellent hydrogen-rich compounds have been designed and predicted to be potential high-temperature superconductors in the past decade[7-9], some of which have been experimentally confirmed. In particular, $H_3S$ and $LaH_{10}$ were first predicted to have superconducting transition temperatures ($T_c$) exceeding 200 K[10-12], which have been confirmed experimentally[13-16]. These are an important milestone in the exploration of hydrogen-based superconductors.

The clathrate hydrides have attracted extensive attention due to their outstanding superconducting. Clathrate hexahydrides *Im-3m*-$XH_6$ (X = Mg, Ca, Sc, Y, La, Tm, Yb, Lu) are widespread in alkaline earth and rare earth metal superhydrides.[12, 17-21]. In this structure, the metal atoms form a body centered cubic (bcc) lattice, and hydrogen atoms occupy all the tetrahedral voids of the bcc lattice, forming a $H_{24}$ cage. $CaH_6$ and $YH_6$ have been experimentally synthesized and exhibit high $T_c$ of 215 K at 172 GPa and 227 K at 166 GPa, respectively[22, 23]. Theoretical predicted $T_c$s of $MgH_6$, $ScH_6$ and $LaH_6$ are 260 K at 300 GPa, 147 K at 285 GPa and 174 K at 100 GPa, respectively. $YbH_6$ and $LuH_6$, with filled f-shelled, are predicted to exhibit high-$T_c$ superconductivity at relatively low pressures (145 K at 70 GPa and 273 K at 100 GPa, respectively)[21]. And $TmH_6$, with unfilled $4f$ orbitals, is stable at 50 GPa, but has a relatively low $T_c$ of 25 K. It is reported that low-pressure stability in lanthanide clathrate hexahydrides correlates strongly with the presence of f states at the Fermi level. With the filling of the f orbitals of the metal atoms, the structure is more easily stabilized at low pressure, but the unfilled f electrons can negatively affect superconductivity.

Incorporating a new element into binary hydrides to form ternary hydrides is an important way to improve the superconducting transition temperature or reduce the superconducting phase stable pressure. In 2019, $Li_2MgH_{16}$ with the highest $T_c$ to date (473 K at 250 GPa), was designed by introducing extra electrons (Li element) to fill the antibonding orbital of the $H_2$ molecular units of $MgH_{16}$[24]. In 2020, a compound of hydrogen, carbon and sulfur showed a superconducting transition at 288 K[25]. However, the stoichiometry and crystal structure of this compound have not yet been determined. This experiment is still subject to many controversies. [26-28]. Although the $T_c$s of $Li_2MgH_{16}$ and C-S-H compound can reach room temperature, the extreme pressure above 250 GPa also makes them difficult for practical application. Recently, a new class of fluorite-type clathrate ternary hydrides $AXH_8$ (A=Ca, Sr, Y, La, X=B, Be, Al) with hydrogen alloy backbone were found [29]. The most outstanding one, $LaBeH_8$, is dynamically stable down to 20 GPa with a high $T_c \sim 185$ K. Although the pressure required for stability of the hydrogen alloy backbone is greatly reduced compared to the pure hydrogen backbone, its $T_c$ is far from room temperature.

The focus of hydrogen-based superconductor research is not simply the pursuit of high-temperature superconductivity or low stable pressure. A good superconductor should achieve a good balance between the pressure required for stability and the critical temperature. The next challenge is to achieve room-temperature superconductivity at significantly low pressures even ambient pressure. Recently, clathrate superhydrides $(La,Y)_{6,10}$ with a maximum critical temperature $T_c$ of 253 K has been experimental synthesized without increasing pressure [30]. This known clathrate hydrides tend to maintain high temperature superconductivity. We note that the heavy rare earth metals Yb/Lu show outstanding properties in sodalite-like clathrate hexahydrides $YbH_6$ and $LuH_6$. With filled f-shells, Yb/Lu can minimize the negative impact on $T_c$ while satisfying the purpose of reducing the pressure.

So, in the present work, we chose heavy rare earth element Yb/Lu doped clathrate hexahydrides to make a step in this goal. A serial of ternary hydrides $A_xB_yH_{6(x+y)}$ (A=Ca, Y, Sc, B=Yb, Lu, and A=Yb, B=Lu, x, y = 1, 2, 3) with high $T_c$ at moderate pressure were found, in which the radii of element A is similar with element B. Among them, the critical temperatures of $Y_3LuH_{24}$, $YLuH_{12}$ and $YLu_3H_{24}$ are 283 K at 120 GPa, 275 K 140 GPa and 288 K at 110 GPa, respectively, which are close to or have reached room temperature. The discovery of $YLu_3H_{24}$ shows that room-temperature superconductivity is possible in hydrogen-based superconductors at moderate pressure.

## 2. Results

We first performed an extensive variable composition structure searches of ternary hydrides Ca-Lu-H, Y-Lu-H and Yb-Lu-H under high pressure. Six sodalite-like clathrate structures $A_xB_yH_{6(x+y)}$ (A=Ca, Y and Yb, B=Lu, x, y = 1, 2, 3) were found in our structure searches, including $Pm$-3$m$ and $Fd$-3$m$ of $ABH_{12}$[31, 32], $P$-3$m$1-$AB_2H_{18}$, $P$-3$m$1-$A_2BH_{18}$, $Fm$-3$m$-$AB_3H_{24}$ and $Fm$-3$m$-$A_3BH_{24}$ (see Fig 1). The thermodynamic stability of these structures was determined by constructing convex hulls (see Fig S1-S4). $Ca_3LuH_{24}$ can be thermodynamically stable at 200 GPa (see Fig S1), and $CaLuH_{12}$, $CaLu_3H_{24}$, $YLu_3H_{24}$, $YLuH_{12}$, $Y_3Lu_2H_{24}$ and $Yb_2LuH_{18}$ are thermodynamically stable at 300 GPa (see Fig S2-4). In addition to hexahydrides, several thermodynamically stable ternary hydrides such as $CaLu_3H_3$, $Ca_3LuH_{15}$, $YLuH_8$, and $Y_3LuH_{20}$ were discovered, which will not be discussed in depth in the subsequent studies (see Fig S1-4).

In order to extend the study to more ternary hydride systems, we substitute the metal elements in these six structures. The properties of clathrate structure can be further improved by choosing a suitable "pre-compressor" element. As discussed above, at least one of them is heavy rare earth element Yb/Lu, and the other element has similar radii with Yb/Lu, including K, Mg, Ca, Sr, Sc, Y, La. We calculated the phonon dispersion for all possible components in the pressure range of 50-200 GPa and finally determined that 36 sodalite-like clathrate hexahydrides can be dynamically stable in seven ternary hydride systems, including Ca-Lu-H, Ca-Yb-H,

Y-Lu-H, Y-Yb-H, Sc-Lu-H, Sc-Yb-H and Yb-Lu-H (see Fig. S9-15). To determine the thermodynamically stable of these structures in Ca-Yb-H, Y-Yb-H, Sc-Lu-H and Sc-Yb-H system, we also performed the structure searches of $A_xB_yH_{6(x+y)}$ and constructed the convex hull, shown in Fig. S5-8. It is reported that the $CaH_6$, $YH_6$ and $YbH_6$ are always thermodynamically stable in the pressure range of 150–300 GPa. Therefore, the energetic stabilities of the Yb-containing hexahydrides in Ca-Yb-H, Y-Yb-H system are evaluated using their formation enthalpies (ΔH) respect to binary hexahydrides. Compared with Lu-containing hexahydrides, many Yb-containing hexahydrides can be thermodynamically stable at 200 GPa, including $CaYbH_{12}$, $Ca_3YbH_{24}$, $CaYb_3H_{24}$ and $YYbH_{12}$. All Sc-containing sodalite-like clathrate hexahydrides we studied are metastable phases (see Fig. S7 and S8).

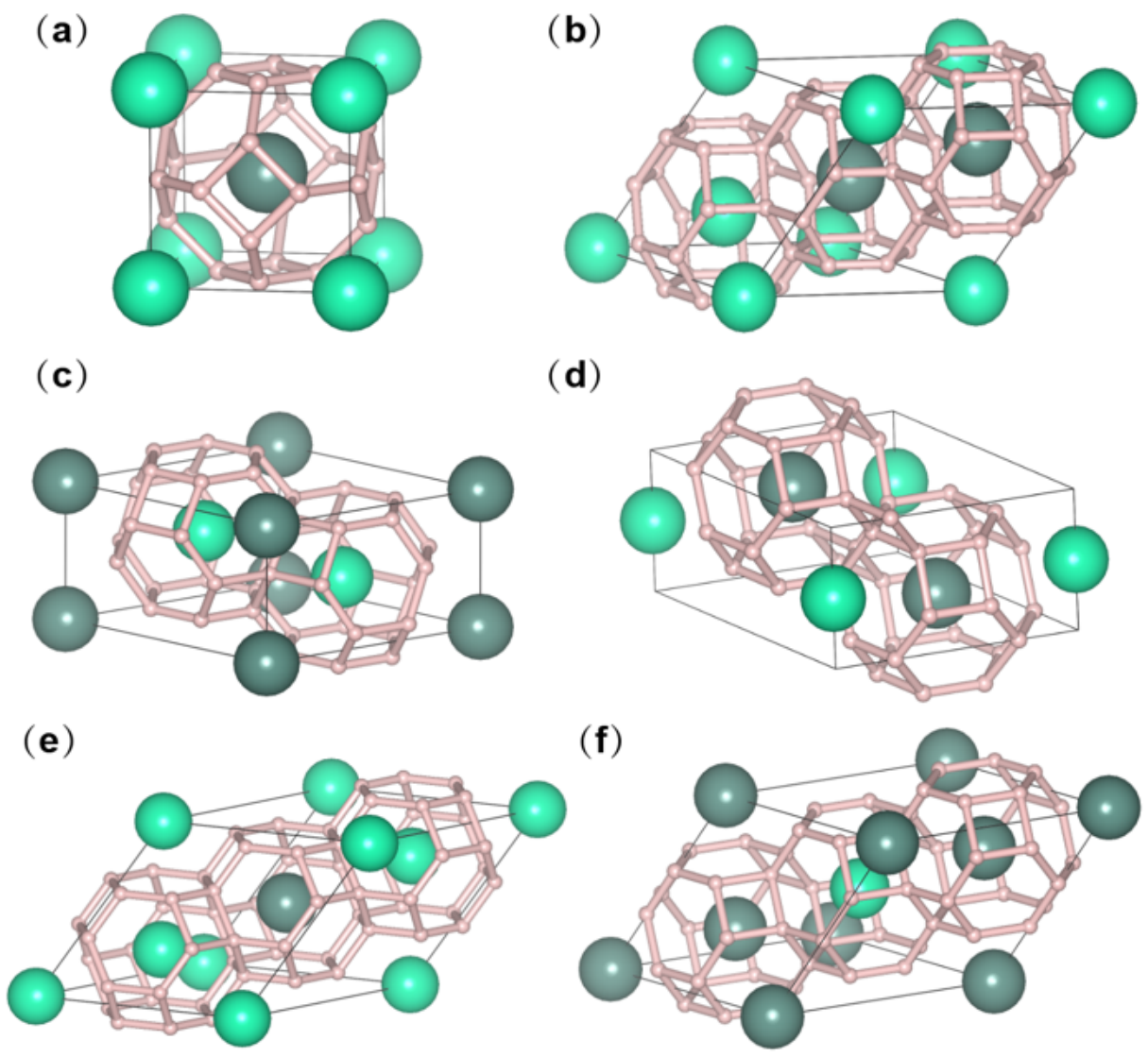

Fig 1. Crystal structures of (a) $Pm$-3$m$-$ABLuH_{12}$, (b) $Fd$-3$m$-$ABH_{12}$, (c) $P$-3$m$1-$AB_2H_{18}$, (d) $P$-3$m$1-$A_2BH_{18}$, (e) $Fm$-3$m$-$AB_3H_{24}$ and (f) $Fm$-3$m$-$A_3BH_{24}$. The light green, dark green and pink atoms represent "pre-compressor" metal atom A and B, respectively. Pink atoms represent H atoms.

After determining the stability of all structures, we further calculated their superconducting properties. The superconductive transition temperatures of these structures are estimated through the Allen–Dynes-modified McMillan equation (A-D-M) with correction factors, and self-consistent solution of the Eliashberg equation (scE) with the Coulomb pseudopotential $\mu^* = 0.10$ and 0.13 (see Table S1). We also calculated the critical temperatures of $CaH_6$ and $YH_6$ in the same way and compared them with the experimentally measured values (see Table S2). The results show that superconducting critical temperature $T_c$ using the self-consistent solution of the Eliashberg equation is in good agreement with the experimental measurements. In addition, the previously proposed metastable phase $Im$-3$m$-$ScH_6$ can be dynamically stable at 130 GPa (see Fig. S13). In this work, we calculated all the properties of $ScH_6$ at 130 GPa.

Figure 2 shows the variation of superconducting critical temperature $T_c$ and minimum dynamically stable pressure with the concentration of heavy rare earth element Lu/Yb in sodalite-like clathrate hexahydrides. Both $T_c$ and minimum pressure of $Ca_{(1-x)}Lu_xH_6$ showed a trend of first increasing and then decreasing with the Lu doping concentration (see Fig 2a). $Ca_3LuH_{24}$ has a $T_c$ of 221 K at 170 GPa, which is close to $CaH_6$. The critical temperature of $CaLuH_{12}$ is 282 K, near room temperature, at 170 GPa. And $CaLu_2H_{18}$ can be dynamically stable at 140 GPa, and its $T_c$ is as high as 299 K at this pressure, which is the highest $T_c$ in this work. It lies only 6 meV/atom above the convex hull at 300 GPa, implying the possibility of experimental synthesis. The images of $T_c$ and minimum pressure of $Y_{(1-x)}Lu_xH_6$ as a function of Lu doping concentration exhibit mirror symmetry (see Fig 2b). $YLu_2H_{18}$ and $Y_2LuH_{18}$ with the *P*-3*m*1 space group have similar properties, with $T_c$ of 242 K and 240 K at 100 GPa, respectively. And $YLu_3H_{24}$ and $Y_3LuH_{24}$ with the *Fm*-3*m* space group also have very close properties, with $T_c$ of 288 K at 110 GPa and 283 K at 120 GPa, respectively. The critical temperature of $YLuH_{12}$ is 275 K at 140 GPa, reaching ice point temperature. For $Y_{(1-x)}Lu_xH_6$, the structure appears to have a greater effect on the $T_c$ and minimum pressure than the doping concentration. For $ScH_6$, doping with Lu can significantly reduce the minimum stable pressure and increase $T_c$. With the increase of Lu doping concentration in $Sc_{(1-x)}Lu_xH_6$, $T_c$ showed an upward trend, and the minimum pressure is basically maintained at about 100 GPa (see Fig 2c), the effect of Lu on reducing the minimum pressure is obvious. The most prominent is $ScLu_3H_{24}$, which reaches a $T_c$ of 271 K at 100 GPa, which is close to ice point temperature. In short, doping Lu in binary hexahydrides achieves the goal of reducing the minimum pressure and increasing $T_c$, such as $CaLuH_{12}$ (282 K at 170 GPa), $CaLu_2H_{18}$ (299 K at 140 GPa), $YLuH_{12}$ (275 K at 140 GPa), $YLu_3H_{24}$ (288 K at 110 GPa) and $Y_3LuH_{24}$ (283 K at 120 GPa).

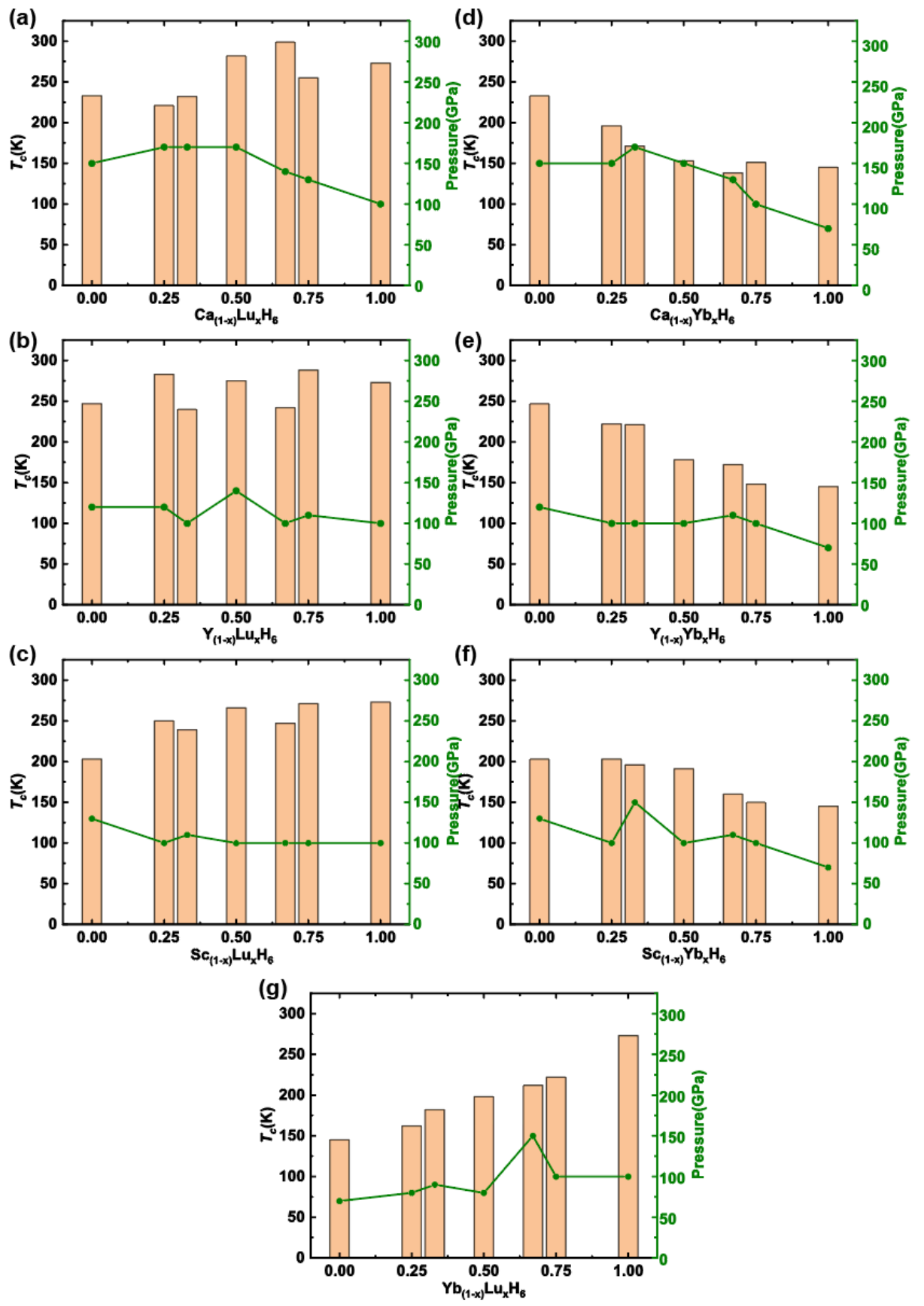

Fig 2. The calculated superconducting critical temperature $T_c$ using the self-consistent solution of the Eliashberg equation at the minmum dynamically stable pressure in (a) $Ca_{(1-x)}Lu_xH_6$, (b) $Y_{(1-x)}Lu_xH_6$, (c) $Sc_{(1-x)}Lu_xH_6$, (d) $Ca_{(1-x)}Yb_xH_6$, (e) $Y_{(1-}$

x)$Yb_xH_6$, (f) $Sc_{(1-x)}Yb_xH_6$ and (g) $Yb_{(1-x)}Lu_xH_6$. The Coulomb pseudopotential is using $\mu^*$ = 0.13.

The $T_c$s of Yb-containing structures are all significantly lower than that of Lu-containing structures, and show a decreasing trend with the increase of Yb doping concentration. Most of the Yb-containing hexahydrides have a $T_c$ of less than 200 K, except for structures with lower concentrations of Yb, such as $Y_3YbH_{24}$ (222 K at 100 GPa), $Y_2YbH_{18}$ (221 K at 100 GPa), $Sc_3YbH_{24}$ (203 K at 100 GPa), $YbLu_2H_{18}$ (212 K at 150 GPa) and $YbLu_3H_{24}$ (222 K at 100 GPa). The effect of Yb on reducing the minimum pressure is obvious. The minimum dynamically stable pressure of $Y_{(1-x)}Yb_xH_6$ is basically maintained around 100 GPa (see Fig 2e). The minimum pressures of $Sc_{(1-x)}Yb_xH_6$ and $Yb_{(1-x)}Lu_xH_6$ decrease with the concentration of Yb doping (See Fig 2f and g). Especially $YbLuH_{12}$ and $Yb_3LuH_{24}$, which contain both heavy rare earth elements Yb and Lu, can be stable at 80 GPa and exhibit $T_c$ of 198 K and 162 K, respectively. They have the lowest stable pressure required in this work. But for $Ca_{(1-x)}Yb_xH_6$, the introduction of heavy rare earth element cannot always reduce the minimum dynamically stable pressure (see Fig 2d). The minimum pressure of $CaYbH_{12}$ and $Ca_3YbH_{24}$ is 150 GPa like that of $CaH_6$, and the minimum pressure of $Ca_2YbH_{18}$ is even higher than that of $CaH_6$, reaching 170 GPa. Only $CaYb_2H_{18}$ and $CaYb_3H_{24}$ with Yb concentrations over 50% exhibited lower pressures than $CaH_6$, with minimum pressures of 130 GPa and 100 GPa, respectively. This is similar to $Ca_{(1-x)}Lu_xH_6$ discussed above. Doping less than 50% Yb/Lu in $CaH_6$ cannot reduce the minimum dynamically stable pressure. This is due to the large difference between the radius of Ca and Yb/Lu.

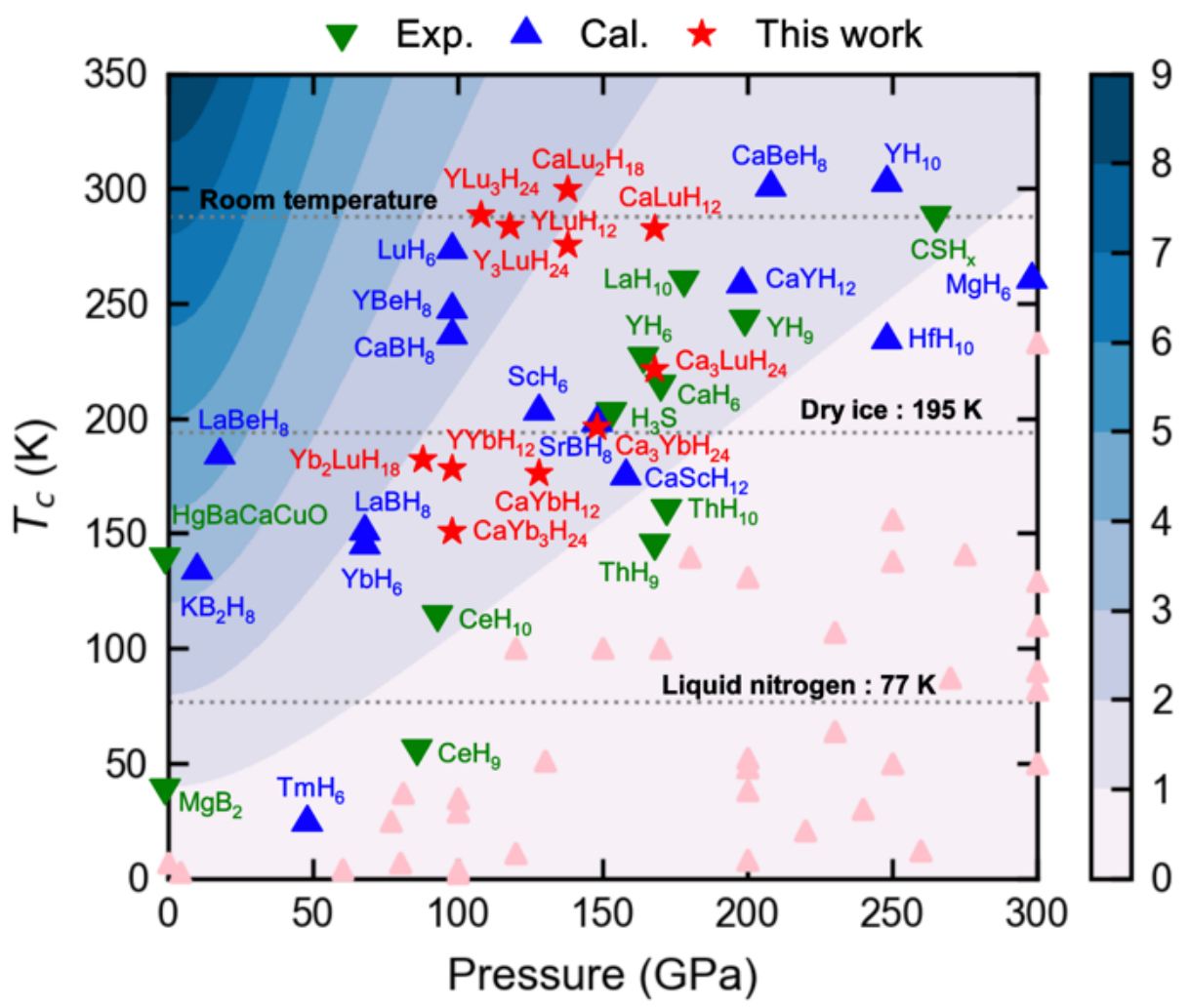

Fig 3. Pressure dependence of $T_c$s calculated for Yb/Lu substituted hexahydrides shown alongside other high-$T_c$ superconductors. The metastable phase $CaLu_2H_{18}$ and all thermodynamically stable phases in this work are marked with red stars. Blue triangles correspond to theoretical predictions[11, 18, 20, 21, 29, 31, 33, 34], green inverted triangles correspond to experimental measurements[13, 16, 22, 23, 25, 35-39], and pink triangles correspond to other less prominent results. The background is shaded according to the figure of merit S.

The focus of hydrogen-based superconductor research is not simply the pursuit of high-temperature superconductivity, after the realization of room-temperature superconductivity. A good superconductor should achieve a good balance between the pressure required for stability and the critical temperature. Therefore, we use a figure of merit S [40] to evaluate the significance of all thermodynamically stable sodalite-like clathrate hexahydrides in this work. S is obtained from the critical temperature $T_c$ and the pressure required for stabilization P:

$$S = \frac{T_c}{\sqrt{T_{c,\mathrm{MgB_2}}^2 + P^2}}$$

$Y_3LuH_{24}$ and $YLu_3H_{24}$ with S > 2, $YLuH_{12}$ and $Yb_2LuH_{18}$ with S close to 2, are better than the well-known hydrogen-based superconductor $H_3S$, $LaH_{10}$, and C-S-H compounds. Most importantly, it can be seen from Fig. 3 that $Y_3LuH_{24}$ is the room temperature superconductor with the lowest pressure required for stability (110 GPa), which is much lower than that of C-S-H, $YH_{10}$ and $CaBeH_8$ with room-temperature superconducting. It also means that room-temperature superconductivity is possible in hydrogen-based superconductors at moderate pressure.

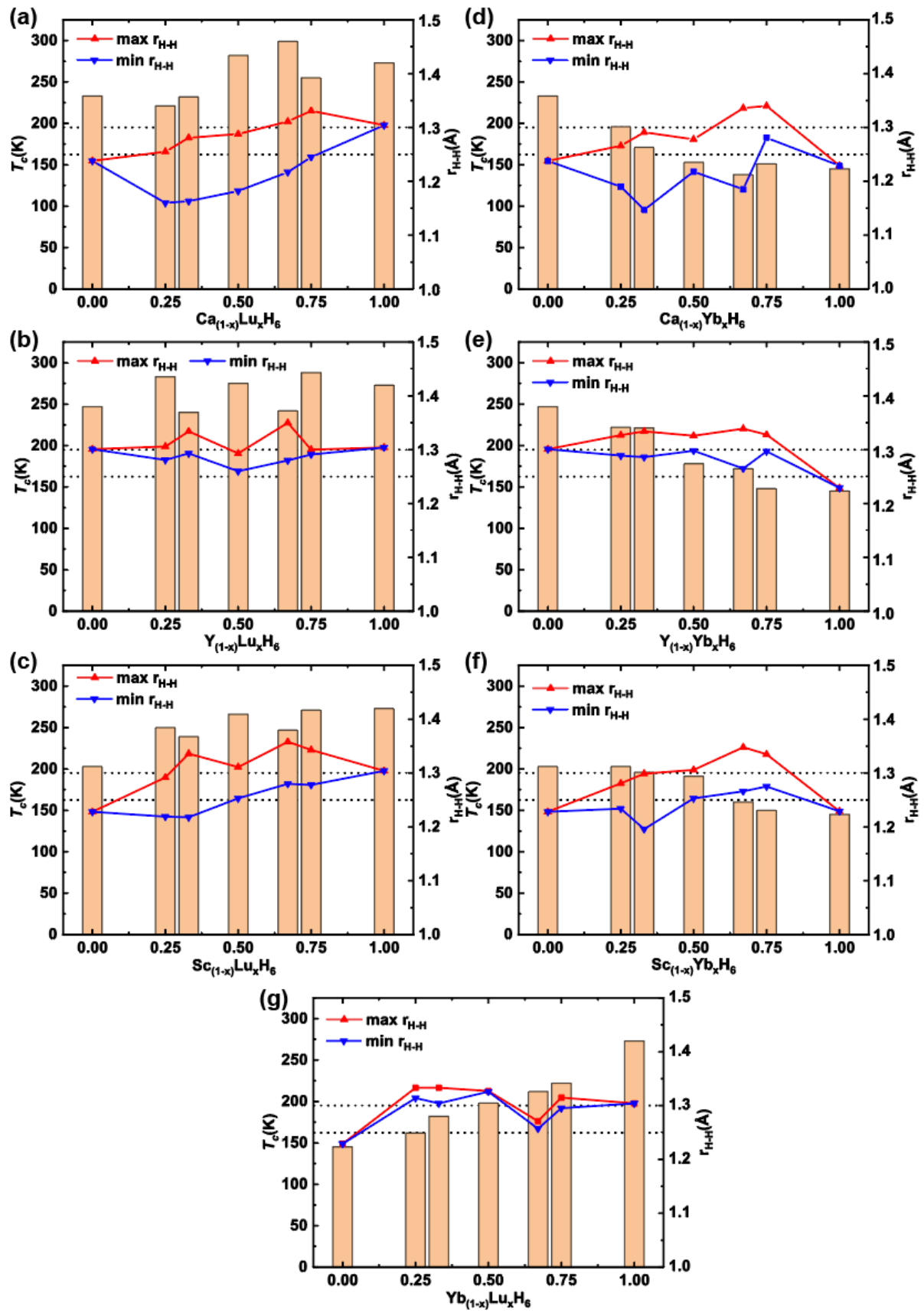

Fig 4. Maximum and minimum H-H bond lengths in (a) $Ca_{(1-x)}Lu_xH_6$, (b) $Y_{(1-x)}Lu_xH_6$, (c) $Sc_{(1-x)}Lu_xH_6$, (d) $Ca_{(1-x)}Yb_xH_6$, (e) $Y_{(1-x)}Yb_xH_6$, (f) $Sc_{(1-x)}Yb_xH_6$ and (g) $Yb_{(1-x)}Lu_xH_6$..

It seems difficult to summarize the law of $T_c$ change only from the perspective of doping concentration for Lu-containing structures. So, we counted the distribution of H-H bond lengths for all hexahydrides in this work. Figure 4 shows the variation of superconducting critical temperature $T_c$ and H-H bond lengths with the concentration of heavy rare earth element Lu/Yb in sodalite-like clathrate hexahydrides. There are nine hexahydrides with $T_c$ over 250 K, including $LuH_6$, $CaLuH_{12}$, $CaLu_2H_{18}$, $CaLu_3H_{24}$, $YLuH_{12}$, $YLu_3H_{24}$, $Y_3LuH_{24}$, $ScLuH_{12}$ and $ScLu_3H_{24}$. They all have one thing in common, that is, the H-H bond lengths are all distributed around 1.25-1.30 Å (see Fig 4a-c). The H-H bond length distribution may be an important factor on $T_c$ for Lu-containing structures. The more H-H bond lengths are distributed around 1.25-1.30 Å, the higher $T_c$ will be. Elongating the H-H bond length to more than 1.25 Å is an effective means to increase $T_c$. In the Yb-containing hexahydrides, $Y_3YbH_{24}$, $Y_2YbH_{18}$, $Sc_3YbH_{24}$, $YbLu_2H_{18}$ and $YbLu_3H_{24}$ have $T_c$ higher than 200 K, in which the H-H bond length distribution is also close to 1.25-1.30 Å (see Fig 4d-g), compared to other Yb-containing hexahydrides. This means that the H-H bond length distribution is likely to be an important reference for finding high-temperature hydrogen-based superconductors in these sodalite-like clathrate hexahydrides. When almost all the H-H bond lengths are in the range of 1.25-1.30 Å, it is very likely to be a potential high-temperature superconductor. Of course, it is limited to these sodalite-like clathrate hexahydrides.

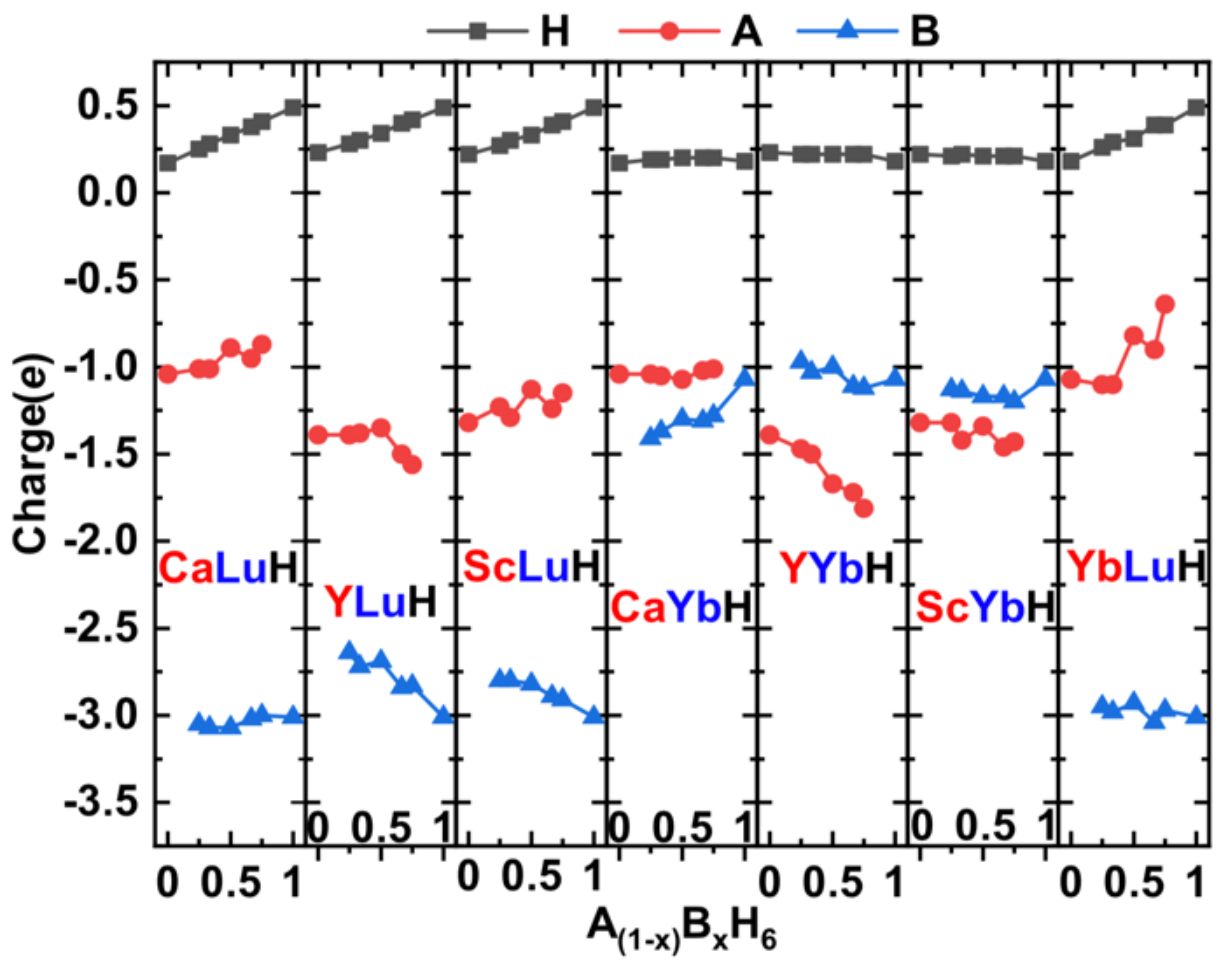

Fig 4. Charges transferred between different elements in $A_{(1-x)}B_xH_6$. Negative mean loss of electrons, positive mean gain of electrons.

Charge transfer is essential for the formation of hydrogen cages in clathrate hydrides. The stability of $H_{24}$ cages in clathrate hexahydrides comes from $H_2$ molecular units accepting electrons from the central metal atoms to form six $H_4$ units as the cornerstone of the construction of the three-dimensional sodalite gabion. Figure 4 shows the charge transfer between two “pre-compressor” metal elements and hydrogen in all sodalite-like clathrate hexahydrides we studied. It can be clearly seen that Lu element is an extremely good electron donor. Each Lu atom can donate 3.07 electrons at most, which is far more than other metal elements. The electrons obtained by H atom in the Lu-containing structures increase with the increasing of the Lu doping concentration (black curves). In the structures without Lu, the number of electrons obtained by H atom is basically the same level (about 0.23 |e|). Although Y atom is also ideal candidates for “pre-compressor” metal element, it can only donate up to 1.81 electrons. Ca, Sc, Yb atoms cannot donate more than 1.5 electrons. Both $YH_6$ and $CaH_6$ exhibit high temperature superconductivity, but they are still far from room temperature superconductor. The introduction of Lu makes it possible to achieve room-temperature superconductivity in sodalite-like clathrate hexahydrides.

## 3. Discussion

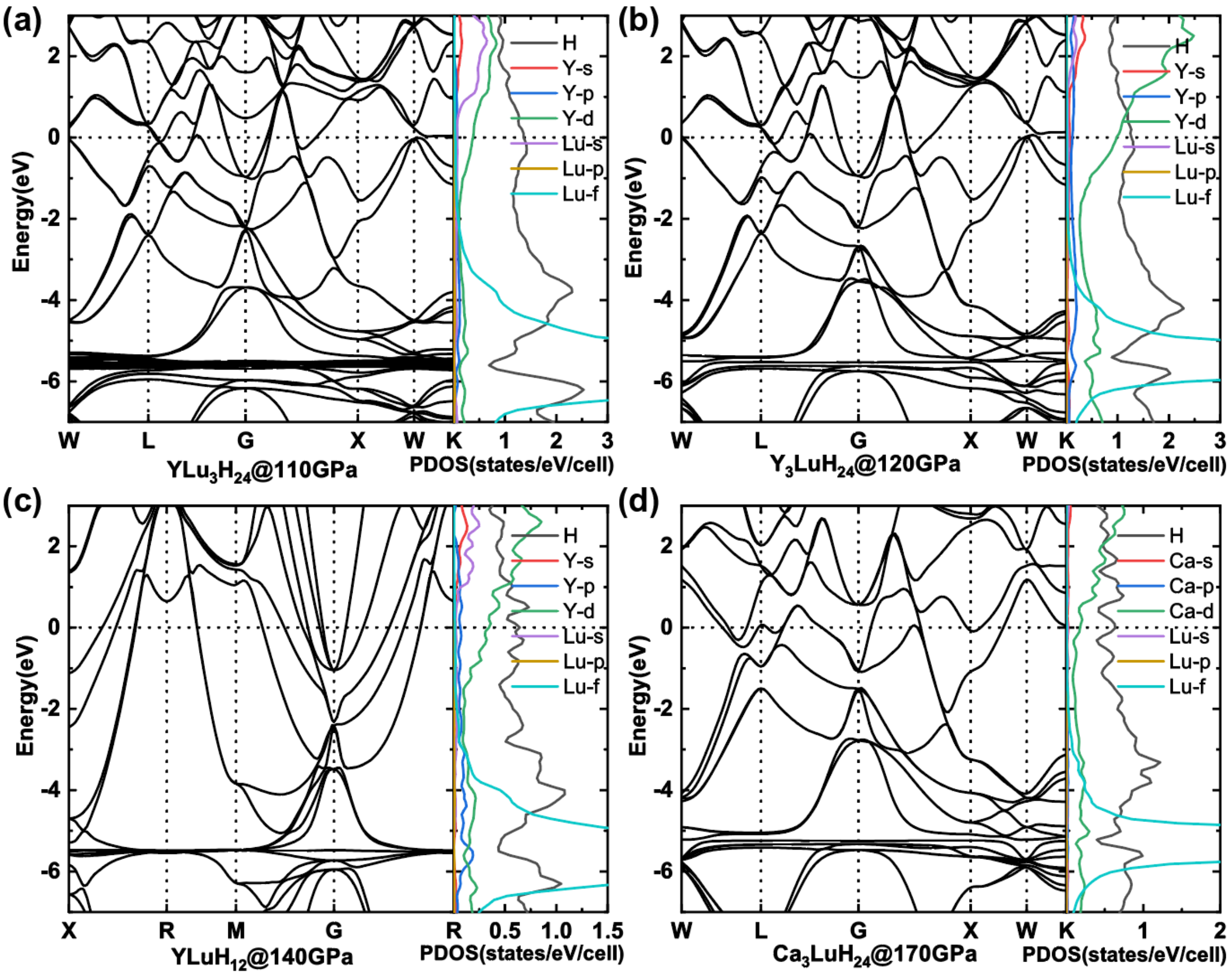

Fig 6. Electronic band structures and projected density of electronic states (PDOS) of (a) *Fm*-3*m*-$YLu_3H_{24}$ at 110 GPa, (b) *Fm*-3*m*-$Y_3LuH_{24}$ at 120 GPa, (c) *Pm*-3*m*-$YLuH_{12}$ at 140 GPa and (d) *Fm*-3*m*-$Ca_3LuH_{24}$ at 170 GPa.

Among Lu-containing clathrate hydrides, the critical temperatures of $Y_3LuH_{24}$, $YLuH_{12}$ and $YLu_3H_{24}$ are 283 K at 120 GPa, 275 K at 140 GPa and 288 K at 110 GPa, respectively, which are close to or have reached room temperature at moderate pressure. In addition, although the $T_c$ of $Ca_3LuH_{24}$ is only 221 K at 170 GPa, it has a lower thermodynamically stable pressure (below 200 GPa), meaning that it is easier to synthesize experimentally. To gain insight into the origin of room-temperature superconductivity for these sodalite-like clathrate hexahydrides, we calculated their electronic band structures and projected density of electronic states (PDOS). For hydrides, the contribution of electronic states of H to the Fermi level is an important basis for judging whether it is an excellent superconductor. In this work, for two structures of the same space group and the same "pre-compressor" metal elements A and B, their contribution of electronic states of H near the Fermi surface are basically the same. For example, the contribution of electronic states of H near the Fermi surface in $YLu_3H_{24}$ and $Y_3LuH_{24}$ are both 1.5 states/eV/f.u. (see Fig 6a and b), which is the reason for their close $T_c$.

In addition to H, f electrons also have an important impact on superconductivity, especially in Yb-containing structures. Note that the $4f$ orbitals associated with the Yb atom form a set of localized bands that appear about 1 eV below the Fermi level (see Fig. S19-22). As the Yb doping concentration increases, the PDOS peak corresponding to the f electrons increases. The position of the peak is always 1 eV below the Fermi level and does not shift with doping concentration, but the contribution of electronic states of the f electrons at the Fermi surface increases with the increase of doping concentration. The 4f-orbital electrons are good for stabilizing the structure, but excessive f electrons at the Fermi surface will negatively affect superconductivity. Therefore, the $T_c$s of Yb-containing structures in Table S1 mostly does not exceed 200 K. Although lowering the concentration of Yb can increase $T_c$, it also makes Yb lose its role in reducing the pressure required for stabilization. For Lu-containing structures, extra electron in the $5d$ orbitals leads to the $4f$ electrons moving away from the Fermi surface in the band structure. Thus, Lu-containing structures are not negatively affected by the 4f-orbital electrons, and exhibits a rather high $T_c$.

In addition, $YLu_3H_{24}$ can exhibit 288 K room temperature superconductivity is also related to the flat band near the Fermi surface along the W-K direction, as shown in Fig 6a and b. Such flat bands are common in sodalite-like clathrate hexahydrides with $Fm$-3$m$ space group in this work, but not all the flat bands can be near the Fermi surface. The energy bands of hydrides are influenced by the pre-compression element. Compared to $CaH_6$, $YH_6$ has a higher Fermi energy due to more valence electrons[20]. $Ca_3LuH_{24}$ also has the same flat band, but at 1.5 eV on the Fermi surface (see Fig 6d). This results in that the contribution of electronic states of H at the Fermi level in $Ca_3LuH_{24}$ is not as high as that of $YLu_3H_{24}$, which in turn leads to a lower $T_c$ than $YLu_3H_{24}$. Due to the lower Lu content, the flat band of $Y_3LuH_{24}$ is slightly above the Fermi surface compared to $YLu_3H_{24}$ (see Fig. 6b). Among all the "pre-compressor" metals, Y and Lu are more favorable for high-temperature superconductivity of $H_{24}$ cage. In ternary clathrate hexahydrides $A_xB_yH_{6(x+y)}$, by selecting the appropriate "pre-compressor" elements A and B, adjust their proportions, maximizing $T_c$ can be achieved.

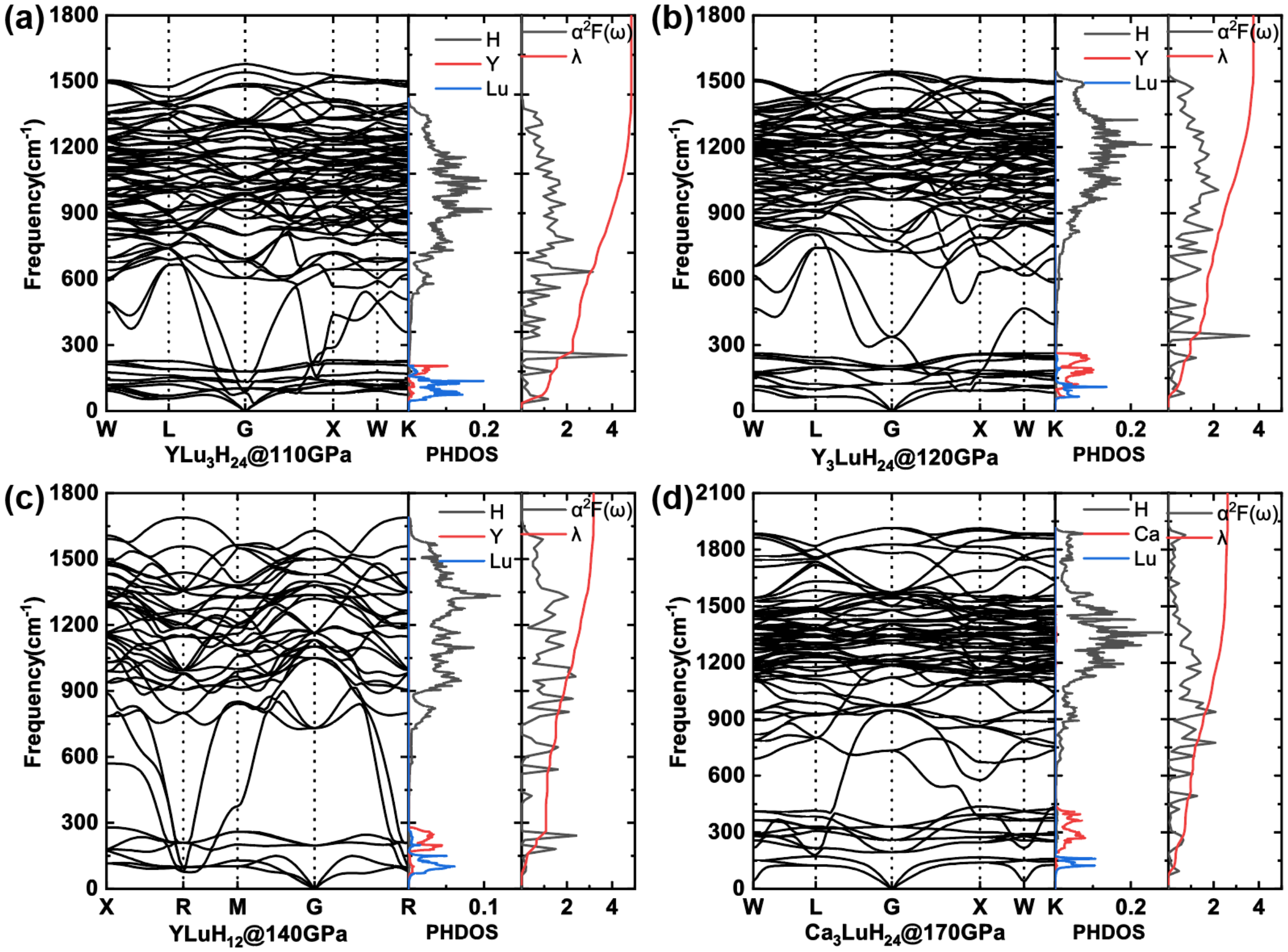

Fig 7. Phonon dispersion, phonon density of state (PHDOS), spectral function $\alpha^2F(\omega)$ and integral EPC parameter $\lambda$ of (a) *Fm*-3*m*-$YLu_3H_{24}$ at 110 GPa, (b) *Fm*-3*m*-$Y_3LuH_{24}$ at 120 GPa, (c) *Pm*-3*m*-$YLuH_{12}$ at 140 GPa and (d) *Fm*-3*m*-$Ca_3LuH_{24}$ at 170 GPa.

The high $T_c$s of hydrogen-based superconductors are largely due to strong electron-phonon coupling (EPC) from high frequency optical phonons. We calculated phonon spectrum, phonon density of state (PHDOS) and integral EPC parameter $\lambda$ of all sodalite-like clathrate hexahydrides we studied, to explore the source of this strong coupling from optical phonons. As shown in Fig 7, the phonon spectrums of these ternary clathrate structures are similar to those of the binary clathrate hexahydrides [17, 19]. The low frequency region is mainly the vibration of metal atoms (red and blue peak in PHDOS), and the high frequency region comes from the vibration of hydrogen (black peak in PHDOS). The maximum vibrational frequency of the hydrogen atom is related to the length of the H-H bond. The shorter the H-H bond, the higher the corresponding vibrational frequency. The maximum vibrational frequency of $H_2$ molecules is generally above 2000 cm$^{-1}$. The absence of spectral lines over 2000 cm$^{-1}$ means that there are no $H_2$ molecules. It can be seen from Fig. 7 that the integral curve of $\lambda$ (red curve) grows rapidly in 500-1500 cm$^{-1}$, while $\lambda$ grows slowly above 1500 cm$^{-1}$ (see Fig 7d). It means that the vibration in frequency range of 500-1500 cm$^{-1}$ is the most important source of electron-phonon coupling. By comparing the phonon spectrum (see Fig. S9-15) and bond length (see Fig 4) of all sodalite-like clathrate hexahydrides, we find that the H-H bond length corresponding to this vibrational frequency range is about 1.25-1.30 Å, which is consistent with the most suitable H-H bond length summarized in Fig. 4. Elongate the H-H bond to 1.25-1.30 Å can reduce the vibrational frequency of hydrogen from high frequency (above 1500 cm$^{-1}$) to a suitable frequency of 500-1500 cm$^{-1}$, thereby increasing $T_c$. In addition to the features shared by clathrate hydrides, these three Y-Lu-H structures have their own unique advantages. It can be seen from Fig. 7a that the integral curve of λ (red curve) of $YLu_3H_{24}$ has reached 3 when the frequency is 500 cm$^{-1}$. The integral curve of λ grows also rapidly in the frequency range 100-500 cm$^{-1}$. On the phonon dispersion, there are some soft phonon modes near the Gamma point in this frequency range of 100-500 cm$^{-1}$. $Y_3LuH_{24}$ and $YLuH_{12}$ also have the same soft phonon modes (see Fig. 5b and c). The integral curve of λ (red curve) corresponding to the frequency at which the phonon softening is also rising rapidly. This indicates that the soft phonon modes play a non-negligible effect on the electron-phonon coupling of optical phonons in $YLu_3H_{24}$ and $Y_3LuH_{24}$ structures.

In conclusion, the incorporation of heavy rare earth elements Yb/Lu is an effective method to tune the superconducting transition temperature and the required pressure for stabilization of sodalite-like clathrate hydrides. In particular, the introduction of Lu element can further improve $T_c$ while keeping the low-pressure stability. The three most prominent compounds $Y_3LuH_{24}$ (283 K at 120 GPa), $YLuH_{12}$ (275 K 140 GPa) and $YLu_3H_{24}$ (288 K at 110 GPa) exhibit room temperature superconductivity at much lower pressure than that of room temperature superconductors, such as C-S-H, $YH_{10}$ and $CaBeH_8$. The enhancement of $T_c$ is achieved by adjusting the H-H bond length to affect the hydrogen vibrational frequency and thereby enhance the electron-phonon coupling. Achieving room-temperature superconductivity at moderate pressure of 100 GPa would represent a major breakthrough in the field of high-temperature superconductivity.

## 4. Computational Methods

High-pressure structure searches were performed using the ab initio random structure searching (AIRSS) technique[41, 42]. For Ca-Lu-H, Y-Lu-H and Yb-Lu-H systems, we predicted more than 8000 structures using variable composition structure searches in each system for their ternary convex hulls. Furthermore, for 36 $A_xB_yH_{6(x+y)}$ (A=Ca, Y, Sc, Yb, B=Yb, Lu, x, y = 1, 2, 3), we predicted about 500 structures for each composition. Structure relaxations during structure searches were performed using the ab initio calculation of the Cambridge Serial Total Energy Package (CASTEP) code[43]. The generalized gradient approximation with the Perdew-Burke-Ernzerhof parametrization[44] for the exchange-correlation functional and ultra-soft pseudo-potentials with cut-off energy of 400 eV and Brillouin zone sampling grid spacing of $2\pi \times 0.07$ Å$^{-1}$ were chose for the structure searching.

Considering the results of pseudo-potential detection for Yb-H and Lu-H in previous work[21], we used CASTEP with ultra-soft pseudo-potentials for structural relaxation and calculations of enthalpies and electronic properties. A cut-off energy of 1000 eV and a Brillouin zone sampling grid spacing [45] of $2\pi\times0.03$ Å$^{-1}$ were used. All enthalpy calculations are well converged to less than 1 meV per atom, which is acceptable for density functional theoretical calculations.

The Quantum-ESPRESSO package[46] was used in phonon and electron–phonon calculations. Ultra-soft pseudo-potentials were used with a kinetic energy cut-off of 90 Ry. The k-points and q-points meshes in the first Brillouin zone of 12×12×12 and 4×4×4 grids were adopted, respectively. The superconductive transition temperatures of these structures are estimated through the Allen−Dynes-modified McMillan equation with correction factors[47, 48] and self-consistent solution of the Eliashberg equation [49].

## Acknowledgments

**Author Contributions:** Defang Duan and Tian Cui initiated the project. Mingyang Du performed the most of the theoretical calculations and contributed to the data interpretation and writing the manuscript. Zihan Zhang and Hao Song contributed to the theoretical calculations. All authors contributed to the discussion and the final version of the manuscript.

**Funding:** This work was supported by National Natural Science Foundation of China (Grants No. 12122405, No. 52072188 and No. 51632002), National Key R&D Program of China (No. 2018YFA0305900), Program for Changjiang Scholars and Innovative Research Team in University (No. IRT_15R23), Parts of calculations were performed in the High Performance Computing Center (HPCC) of Jilin University and TianHe-1 (A) at the National Supercomputer Center in Tianjin.

**Conflicts of Interest**
The authors declare that there is no conflict of interest regarding the publication of this article."

# Supplementary Material

# Room-temperature superconductivity in Yb/Lu substituted sodalite-like clathrate hydrides under moderate pressure

Mingyang Du[1], Hao Song[1], Zihan Zhang[1], Defang Duan[1]*, Tian Cui[1,2]*

[1]*State Key Laboratory of Superhard Materials, College of Physics, Jilin University, Changchun 130012, People's Republic of China*

[2]*School of Physical Science and Technology, Ningbo University, Ningbo, 315211, People's Republic of China*

Correspondence author: *cuitian@jlu.edu.cn, †duandf@jlu.edu.cn

# FIGURES

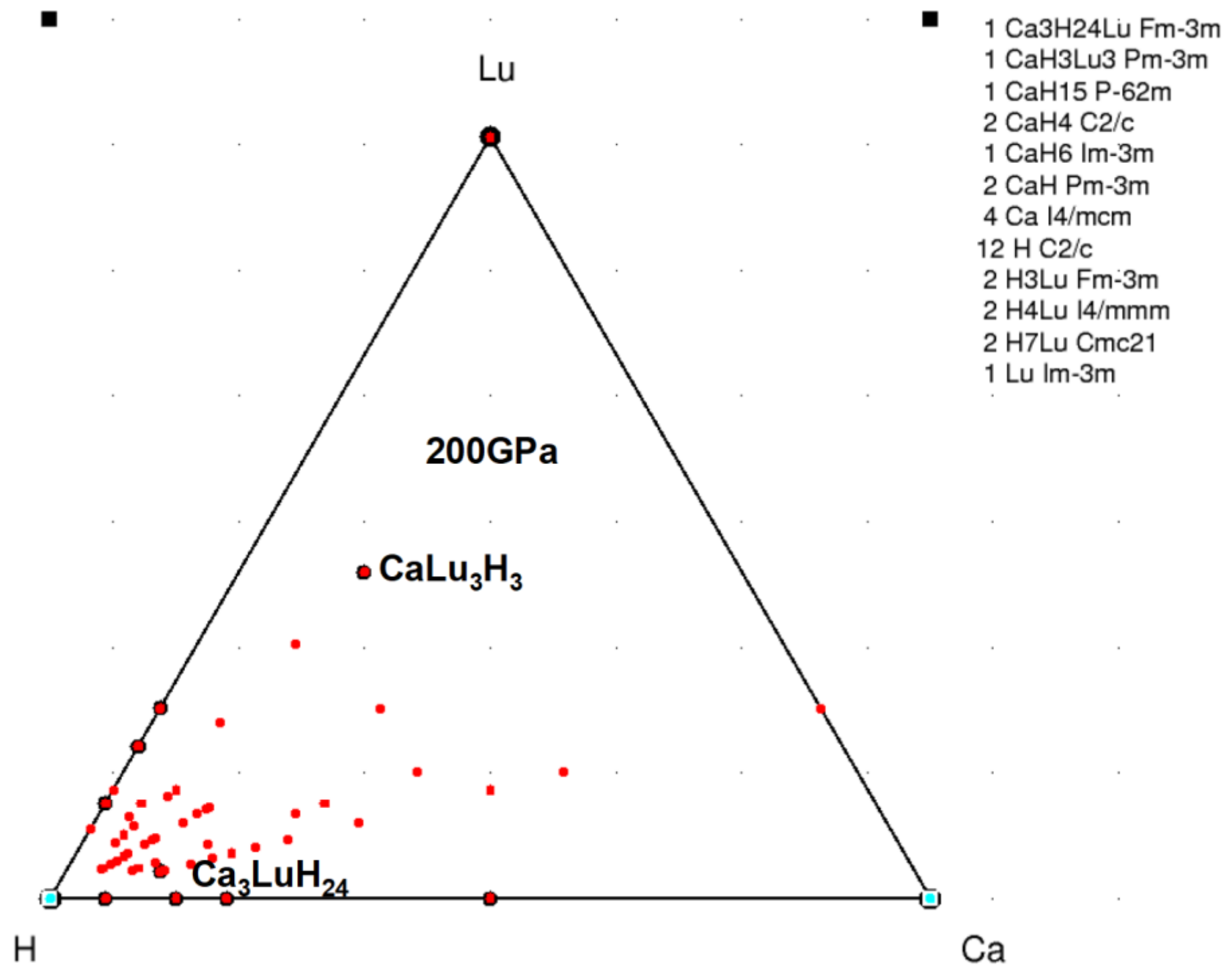

Fig. S1 The convex hull of Ca-Lu-H system at 200 GPa. The corresponding elements and boundary binary phases are chosen from the results of the previous works[1, 2].

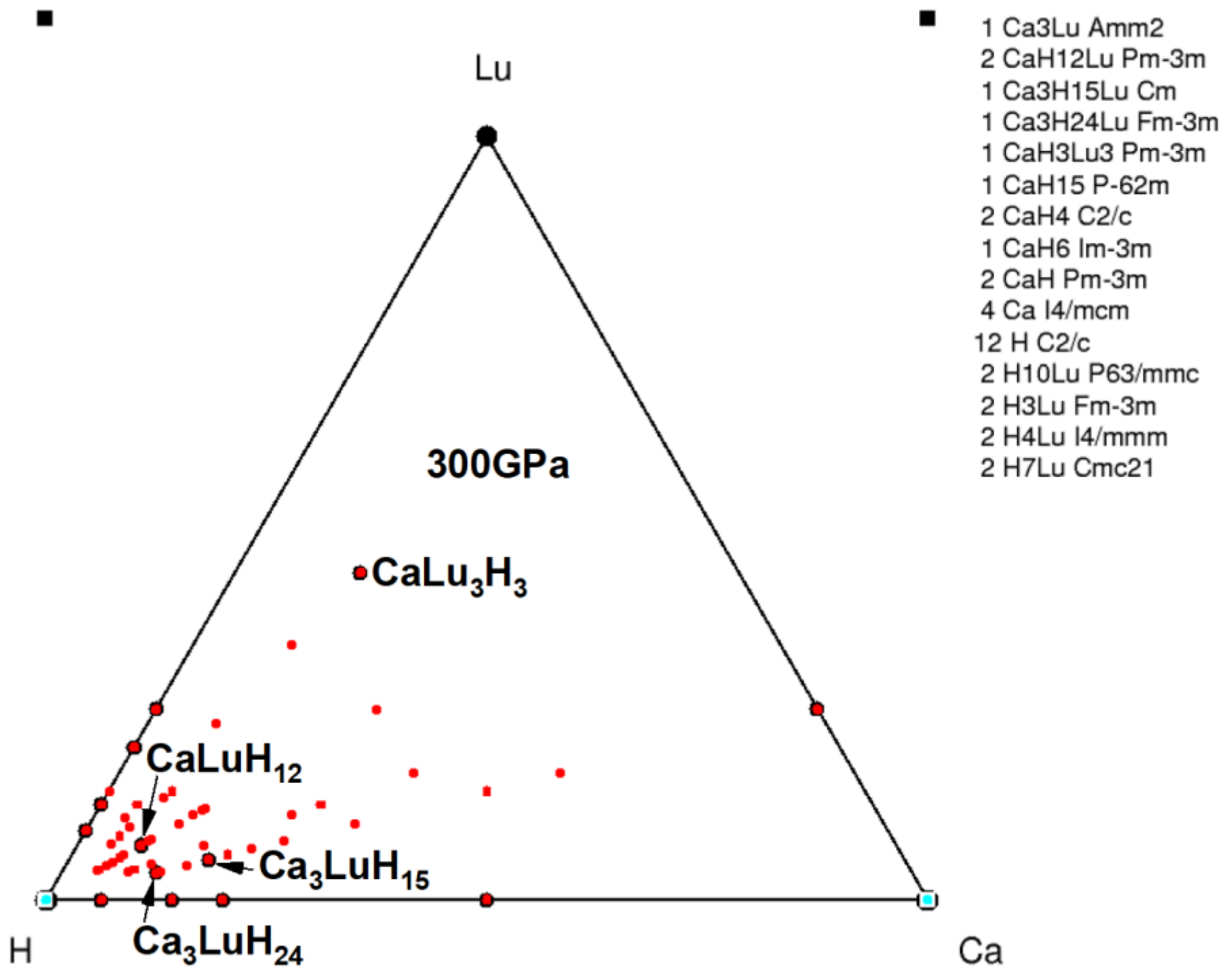

Fig. S2 The convex hull of Ca-Lu-H system at 300 GPa. The corresponding elements and boundary binary phases are chosen from the results of the previous works[1, 2].

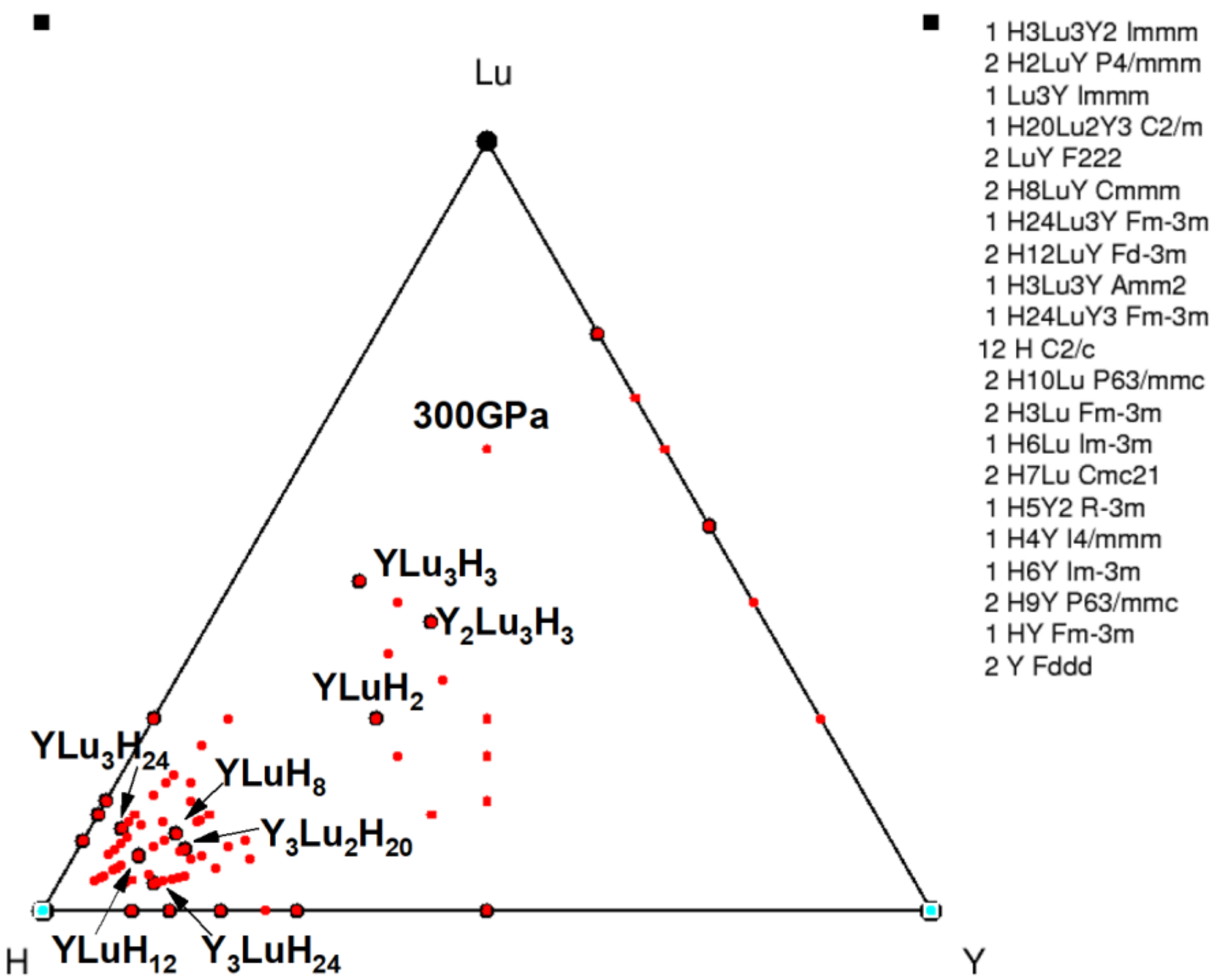

Fig. S3 The convex hull of Y-Lu-H system at 300 GPa. The corresponding elements and boundary binary phases are chosen from the results of the previous works[2-4].

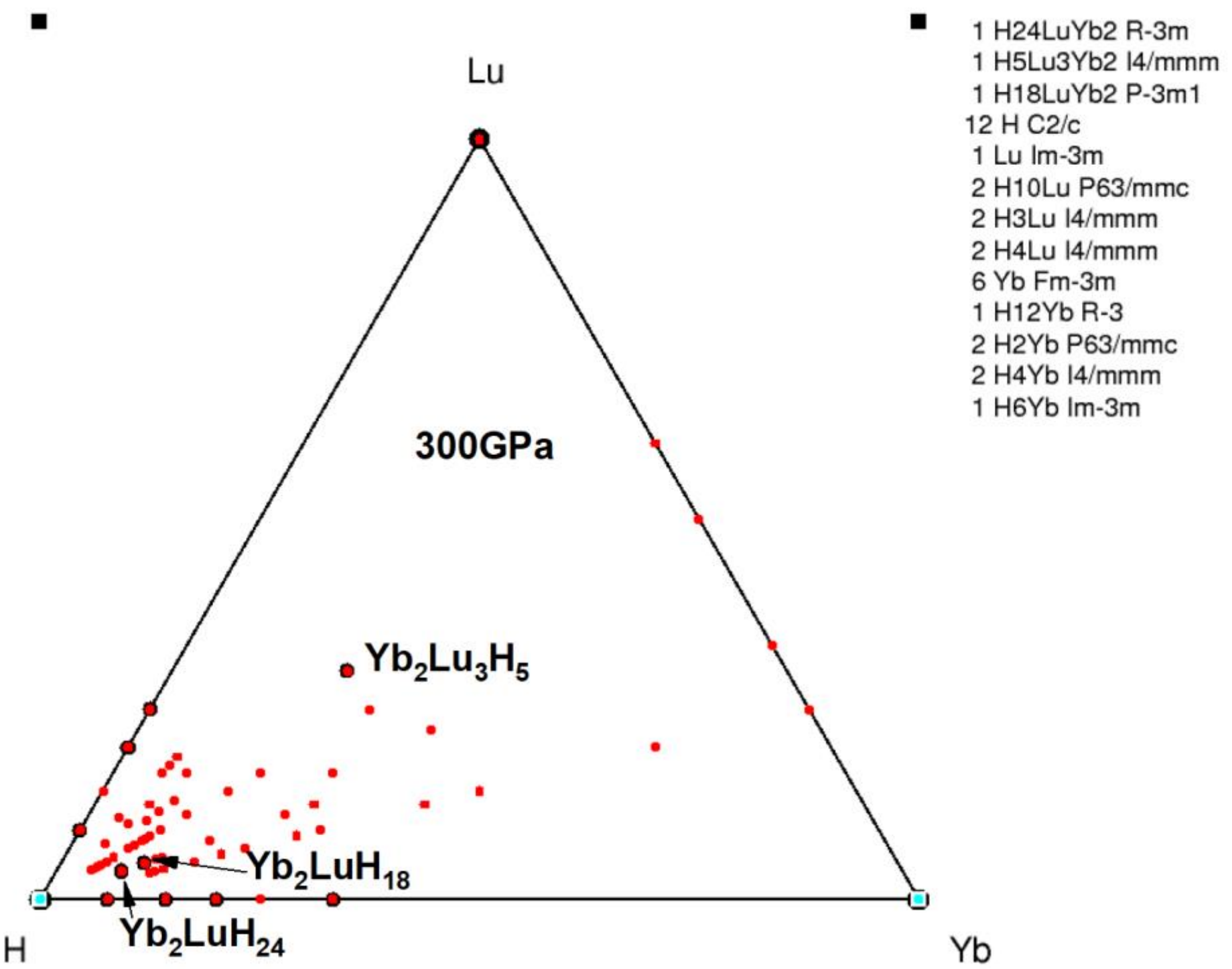

Fig. S4 The convex hull of Yb-Lu-H system at 300 GPa. The corresponding elements and boundary binary phases are chosen from the results of the previous works[2].

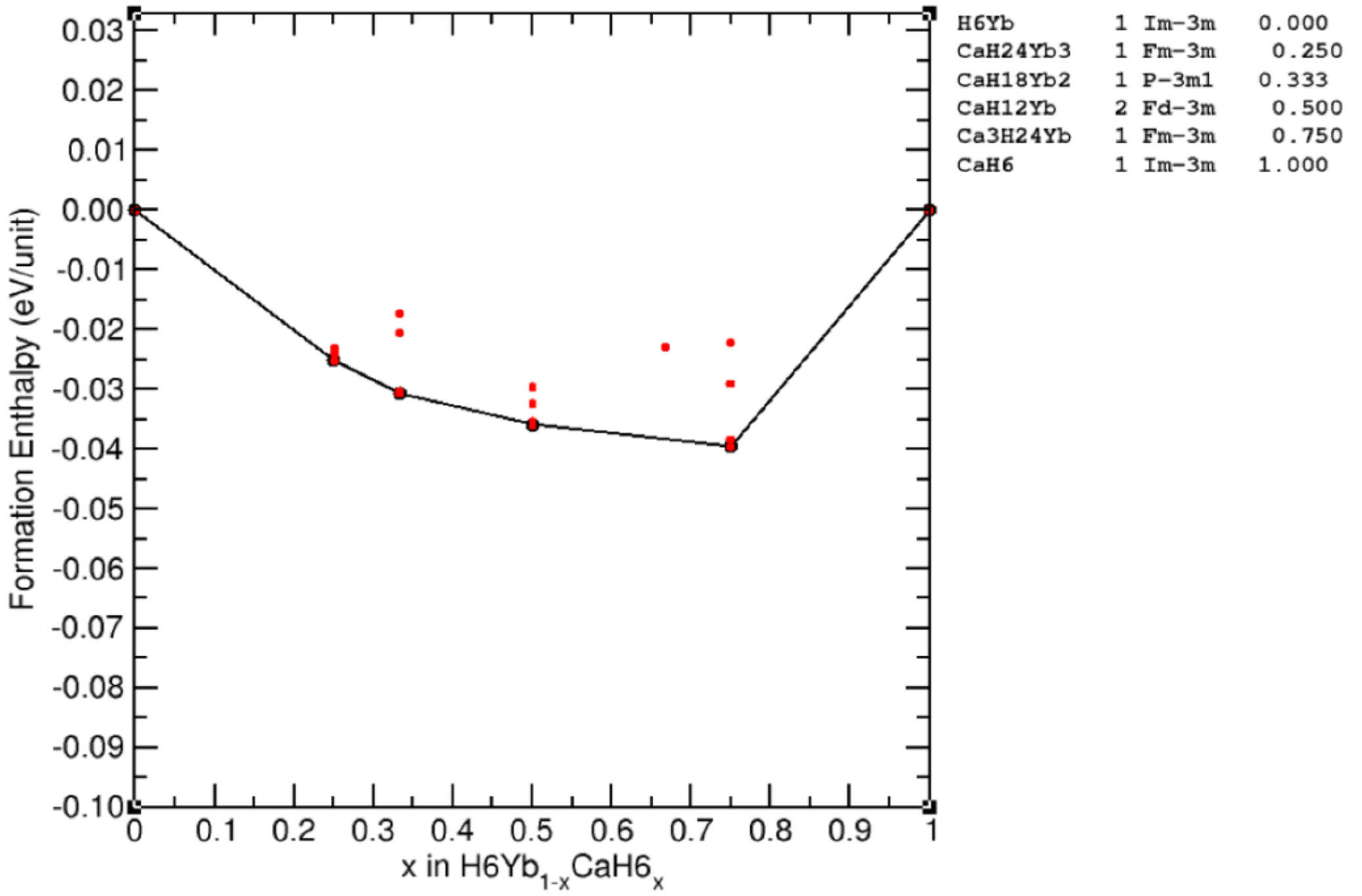

Fig. S5 The convex hull of Ca-Yb-H system at 200 GPa. The corresponding boundary binary phases are chosen from the results of the previous works[1, 2].

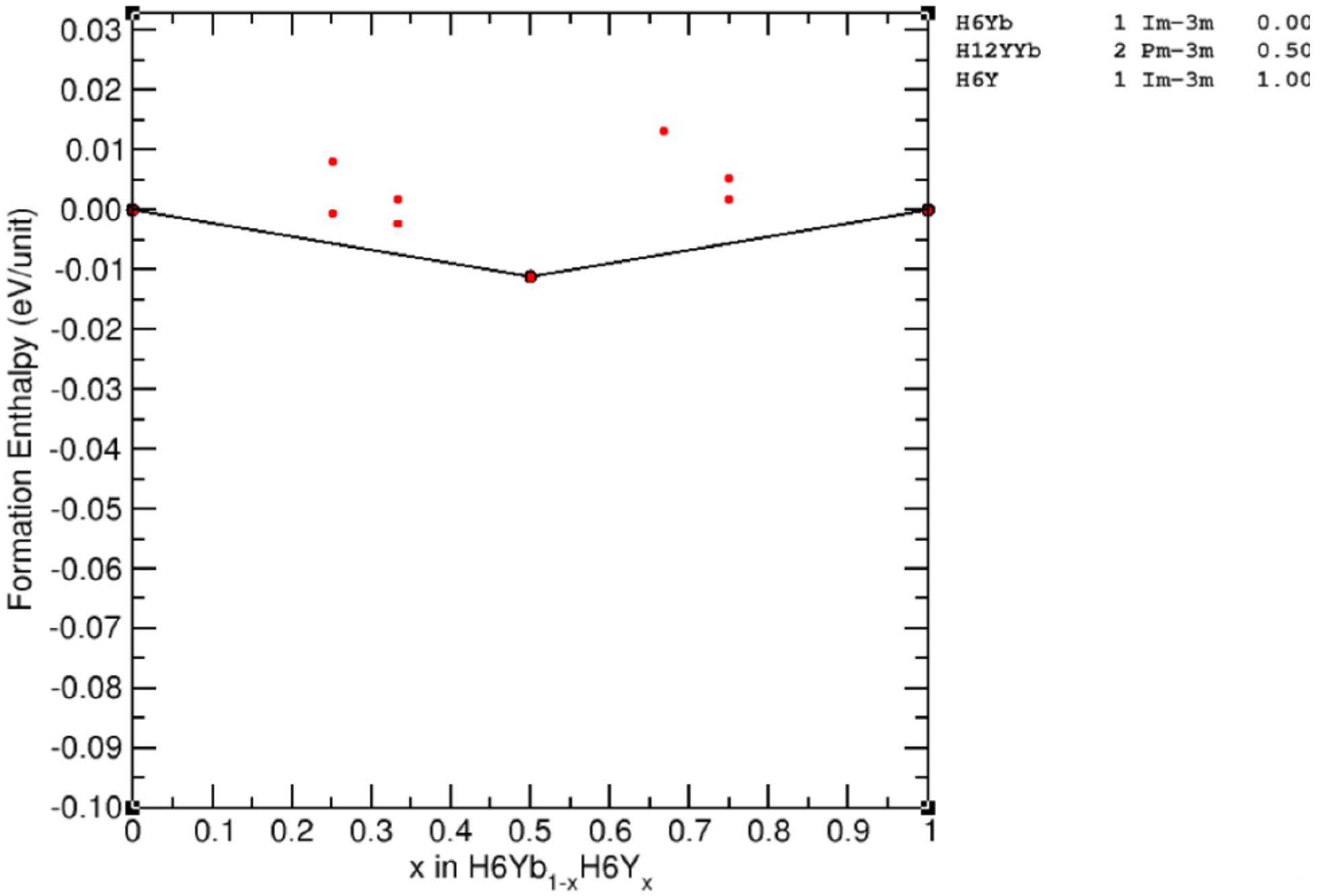

Fig. S6 The convex hull of Y-Yb-H system at 200 GPa. The corresponding boundary binary phases are chosen from the results of the previous works[2, 3].

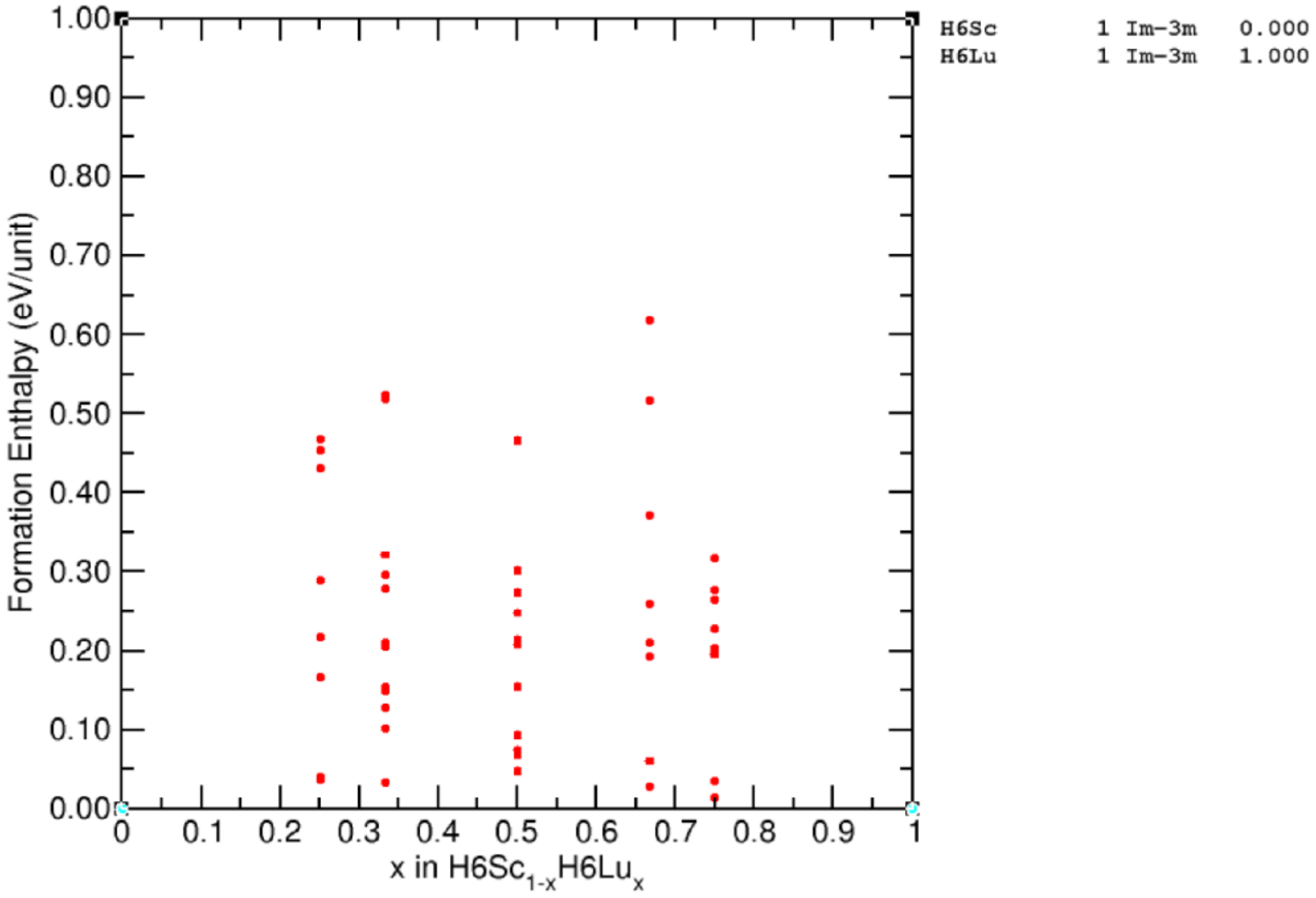

Fig. S7 The convex hull of Sc-Lu-H system at 300 GPa. The corresponding boundary binary phases are chosen from the results of the previous works [2, 5].

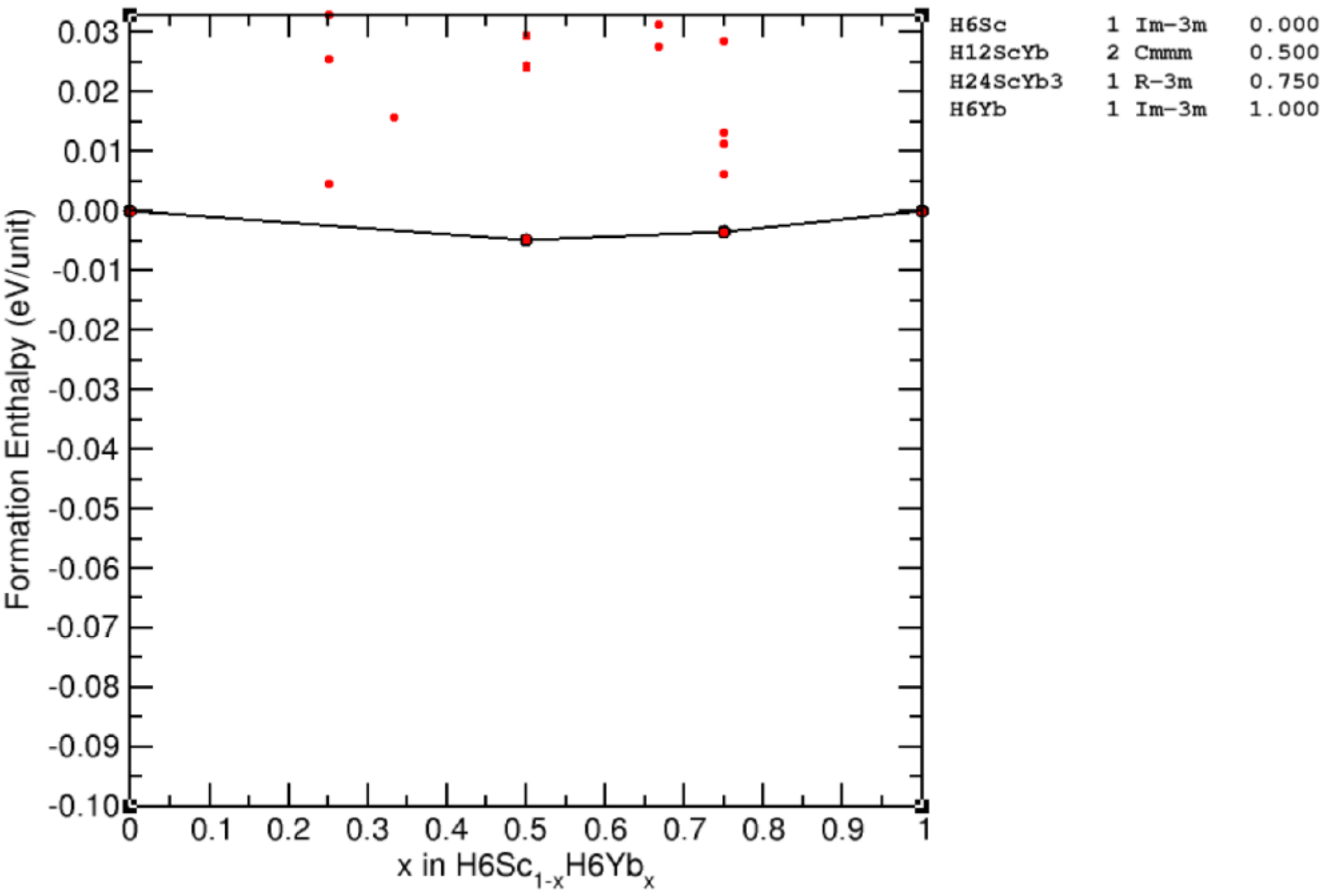

Fig. S8 The convex hull of Sc-Yb-H system at 300 GPa. The corresponding boundary binary phases are chosen from the results of the previous works [2, 5].

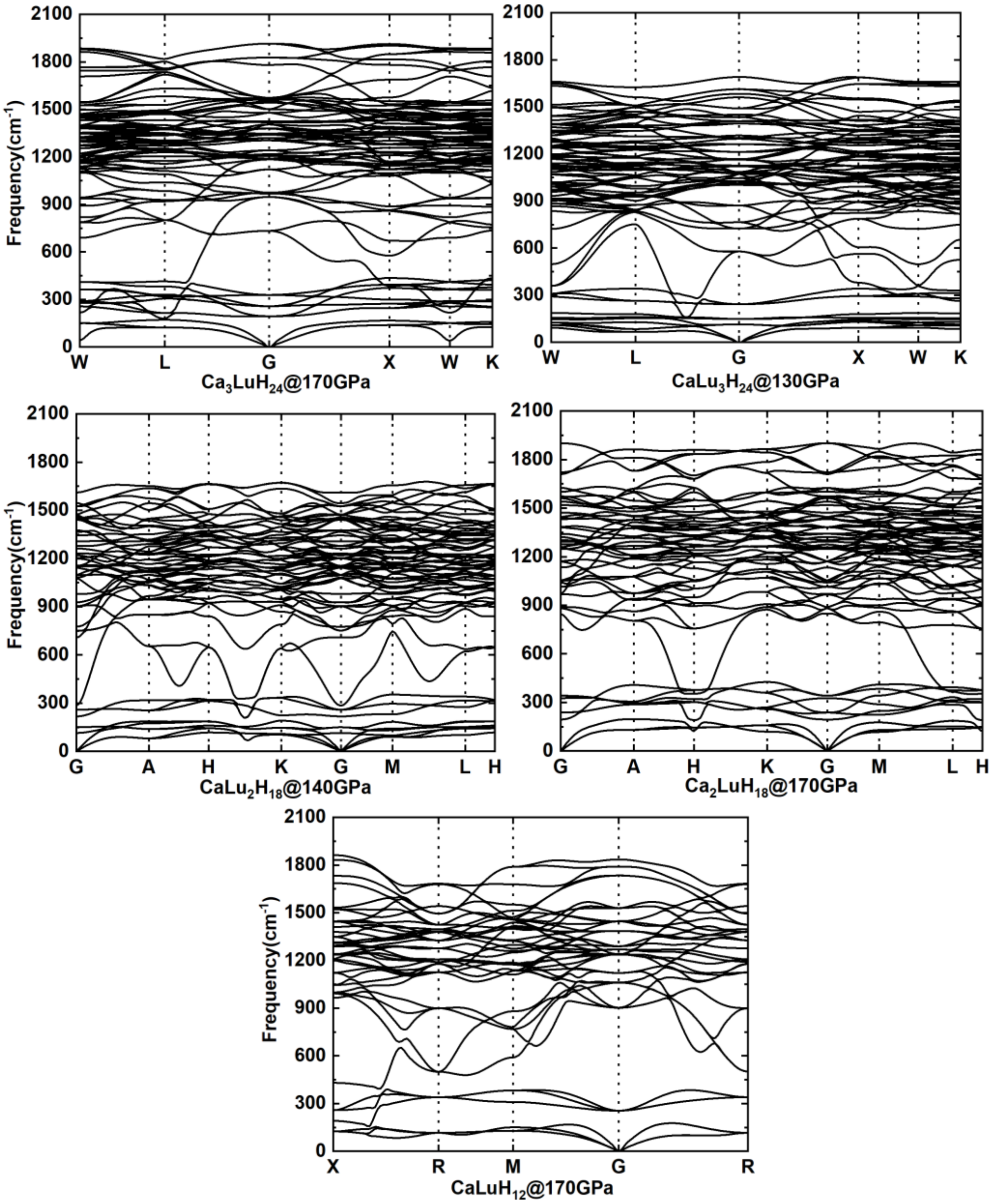

Fig. S9 The phonon band structure of $Ca_3LuH_{24}$, $CaLu_3H_{24}$, $CaLu_2H_{18}$, $Ca_2LuH_{18}$, $Ca_3LuH_{24}$ and $CaLuH_{12}$ under their minimum dynamically stable pressures, respectively.

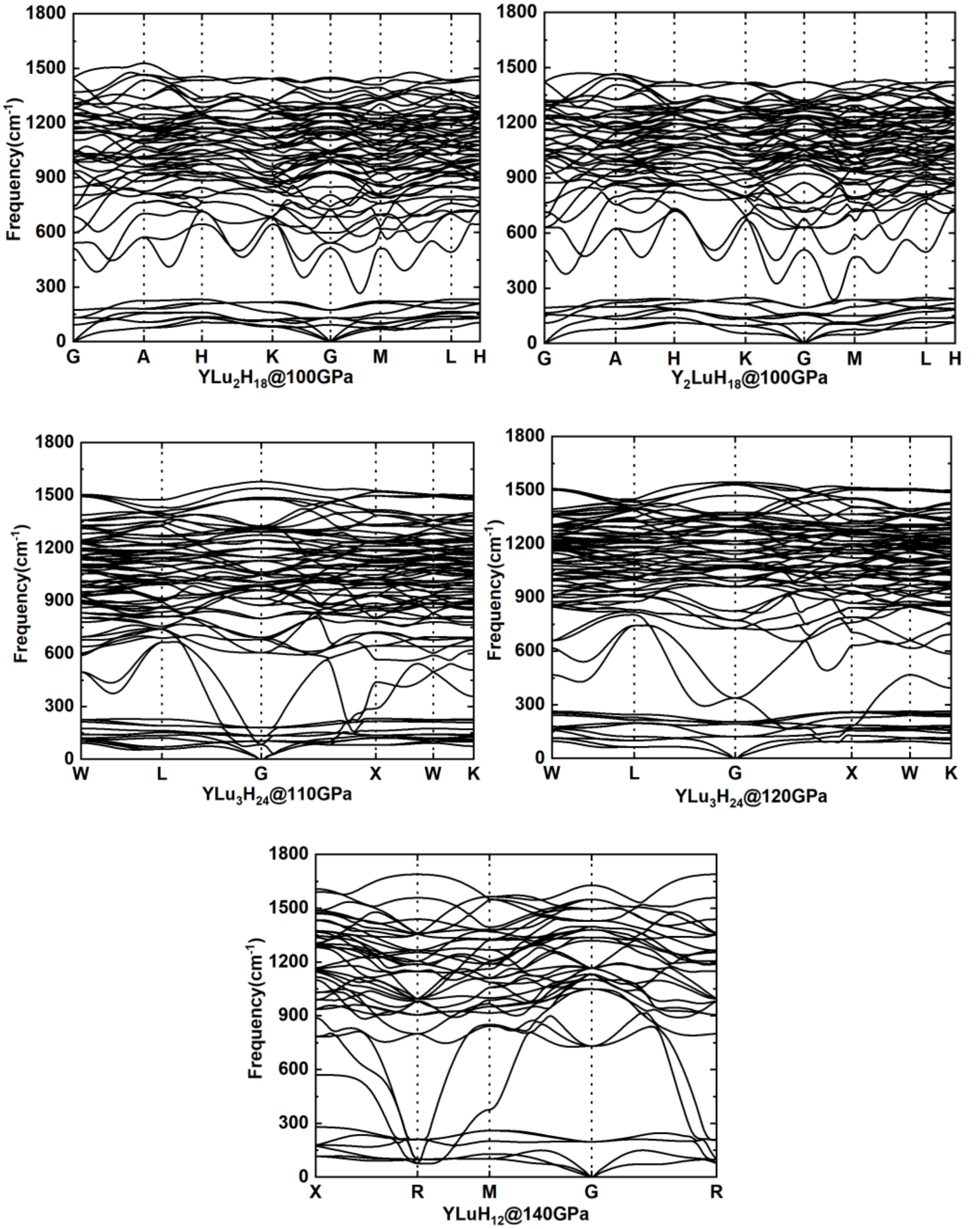

Fig. S10 The phonon band structure of $YLu_2H_{18}$, $Y_2LuH_{18}$, $YLu_3H_{24}$, $Y_3LuH_{24}$ and $YLuH_{12}$ under their minimum dynamically stable pressures, respectively.

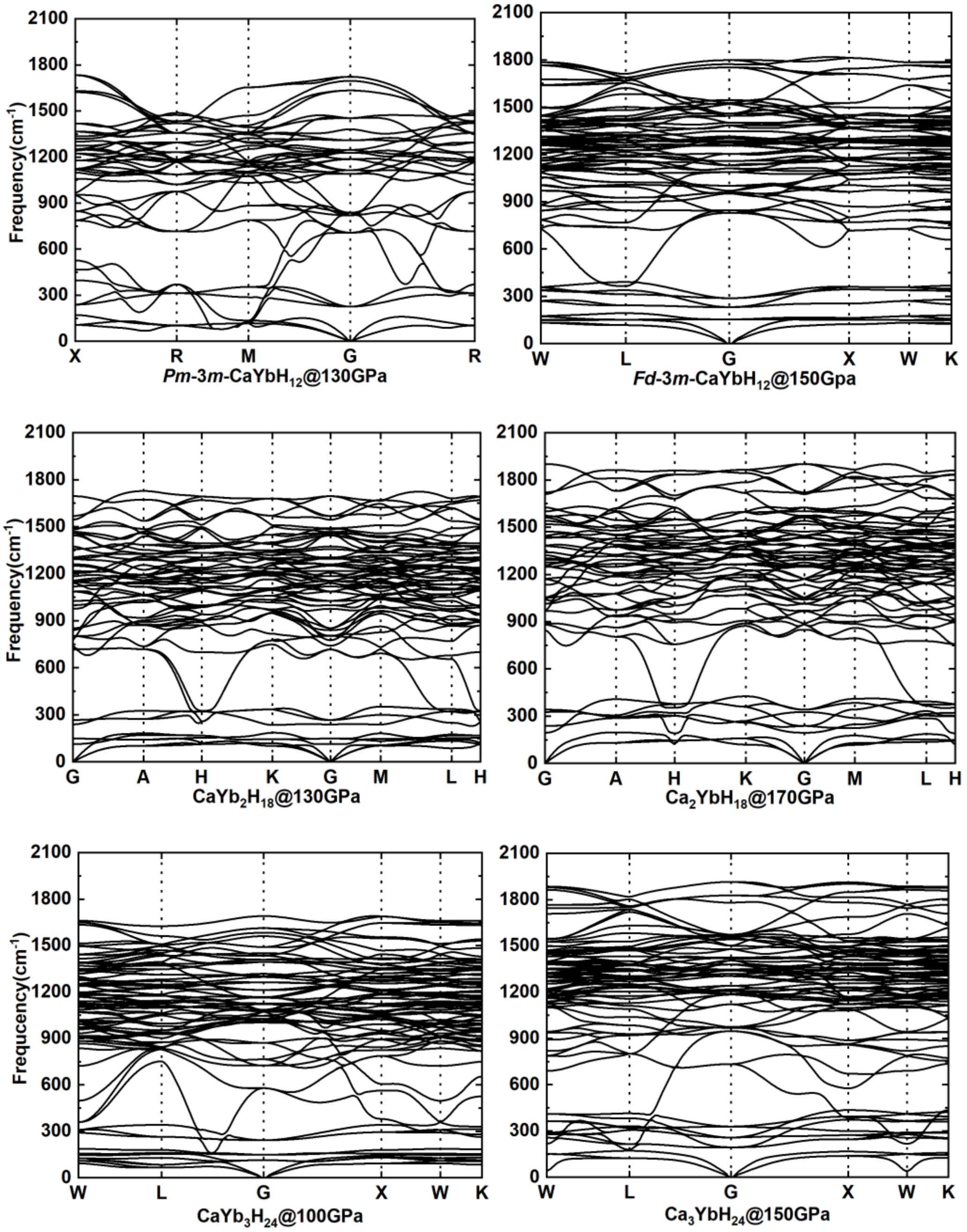

Fig. S11 The phonon band structure of *Pm*-3*m*-$CaYbH_{12}$, *Fd*-3*m*-$CaYbH_{12}$, $CaYb_2H_{18}$, $Ca_2YbH_{18}$, $CaYb_3H_{24}$ and $Ca_3YbH_{24}$ under their minimum dynamically stable pressures, respectively.

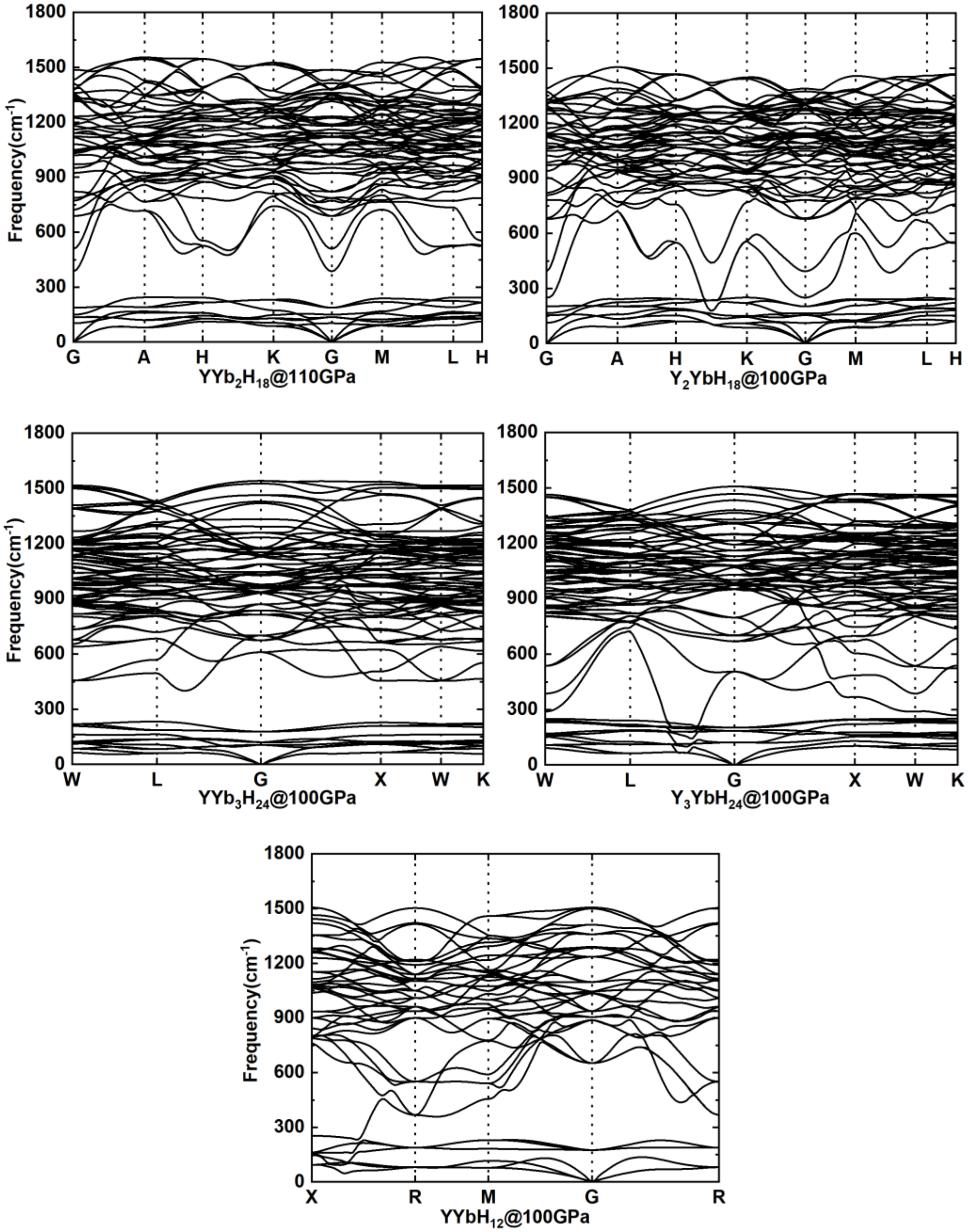

Fig. S12 The phonon band structure of $YYbH_{12}$, $YYb_2H_{18}$, $Y_2YbH_{18}$, $YYb_3H_{24}$ and $Y_3YbH_{24}$ under their minimum dynamically stable pressures, respectively.

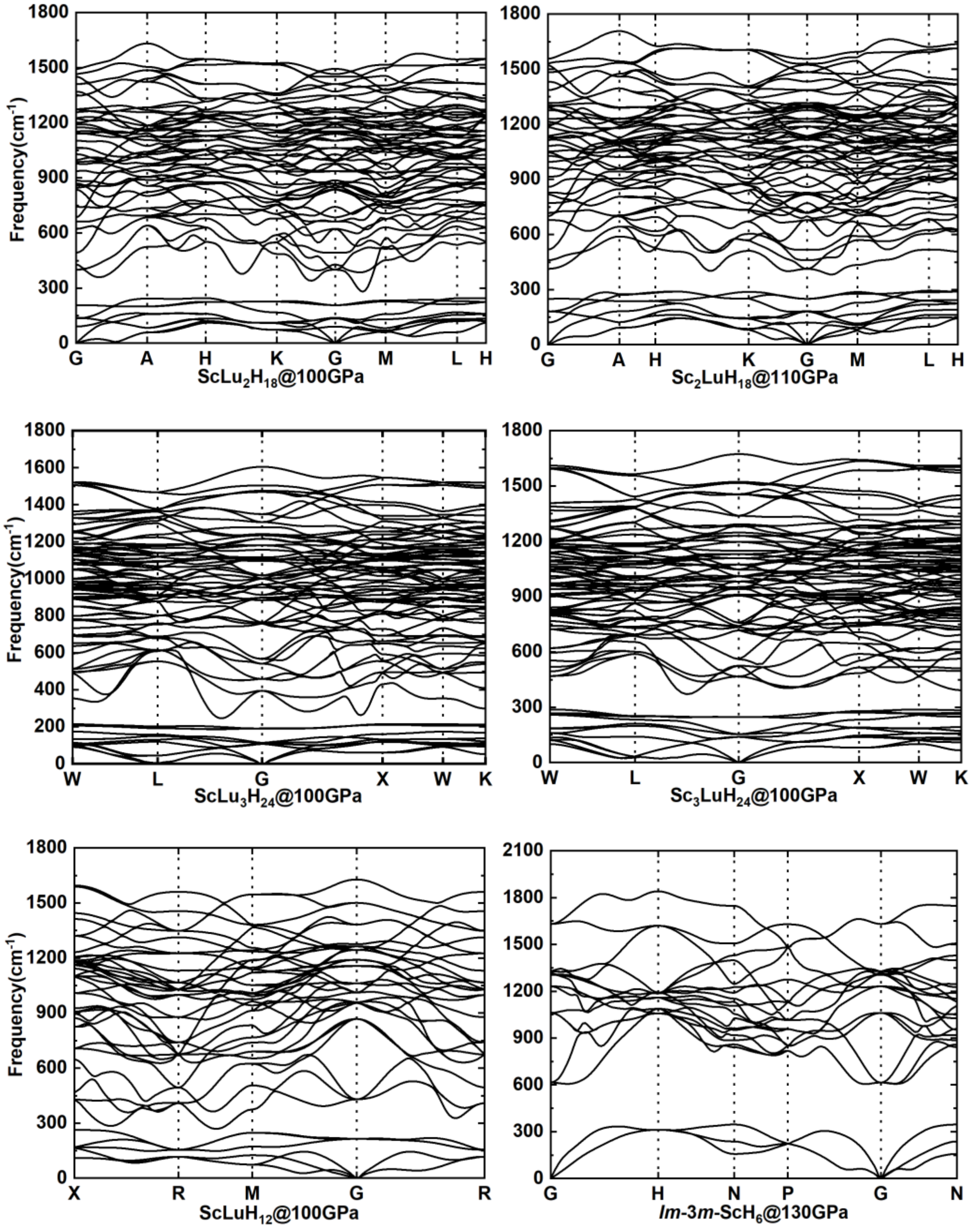

Fig. S13 The phonon band structure of $ScLu_2H_{18}$, $Sc_2LuH_{18}$, $ScLu_3H_{24}$, $Sc_3LuH_{24}$ $ScLuH_{12}$, and $ScH_6$ under their minimum dynamically stable pressures, respectively.

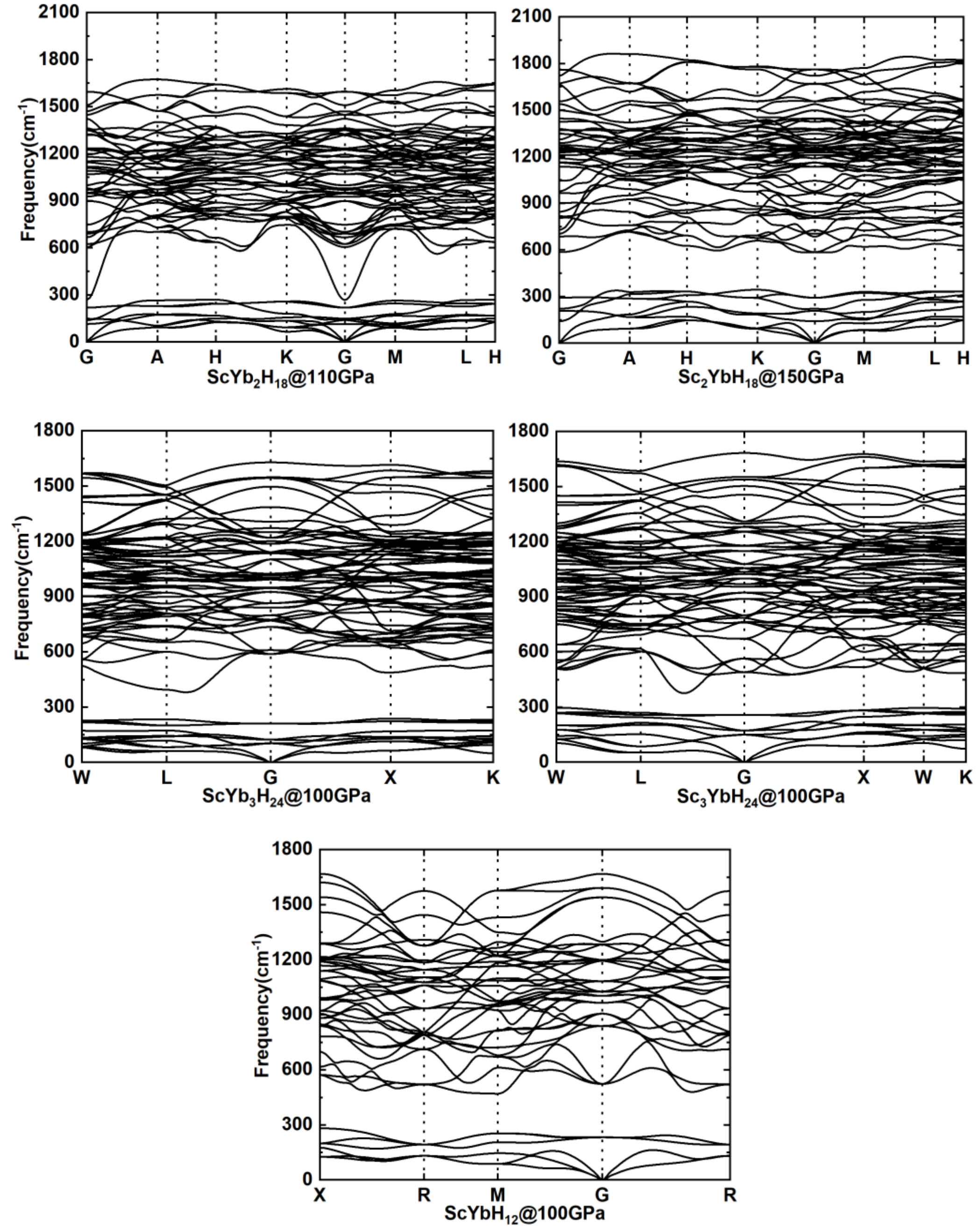

Fig. S14 The phonon band structure of $ScYb_2H_{18}$, $Sc_2YbH_{18}$, $ScYb_3H_{24}$, $Sc_3YbH_{24}$ and $ScYbH_{12}$ under their minimum dynamically stable pressures, respectively.

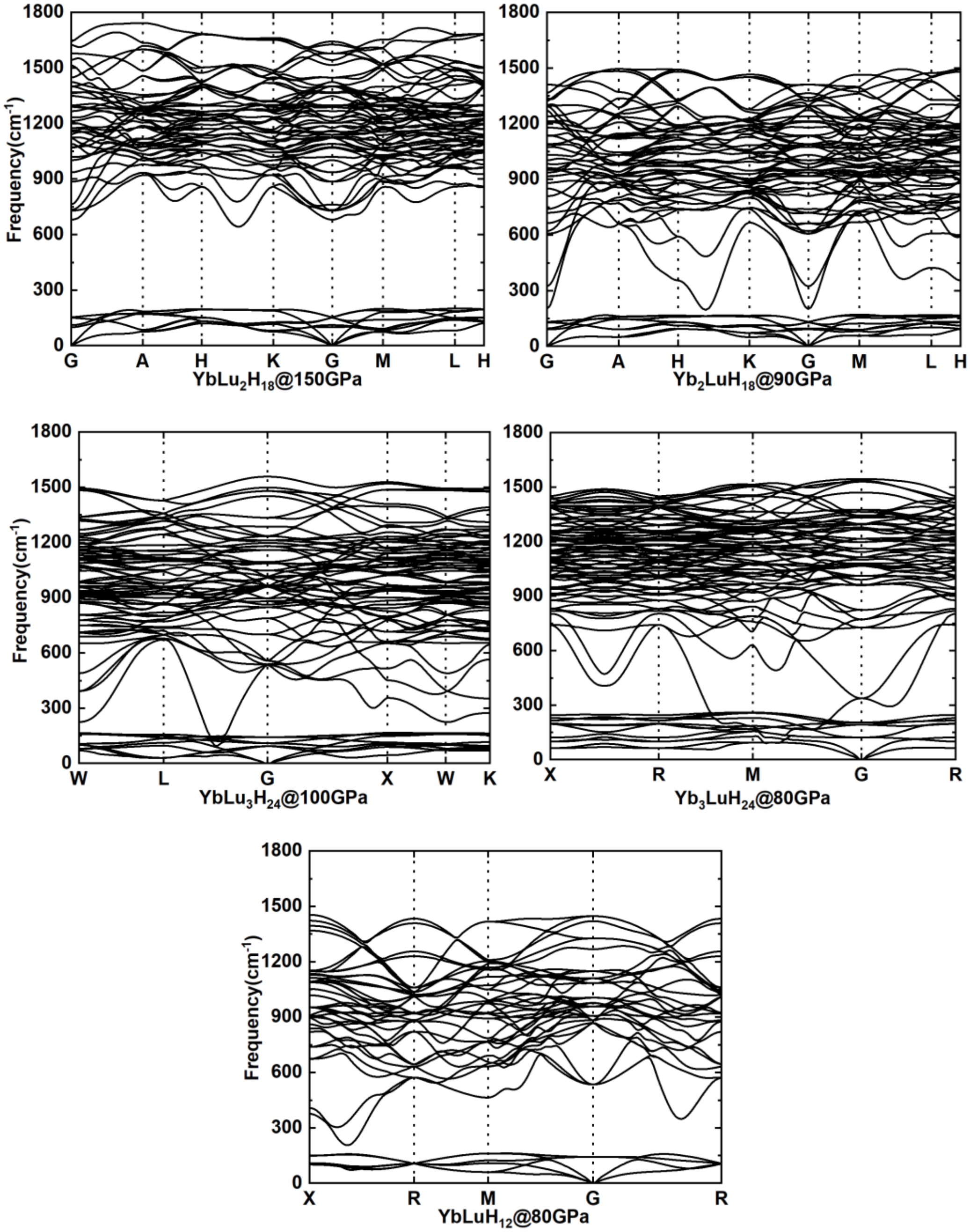

Fig. S15 The phonon band structure of $YbLu_2H_{18}$, $Yb_2LuH_{18}$, $YbLu_3H_{24}$, $Yb_3LuH_{24}$ and $YbLuH_{12}$ under their minimum dynamically stable pressures, respectively.

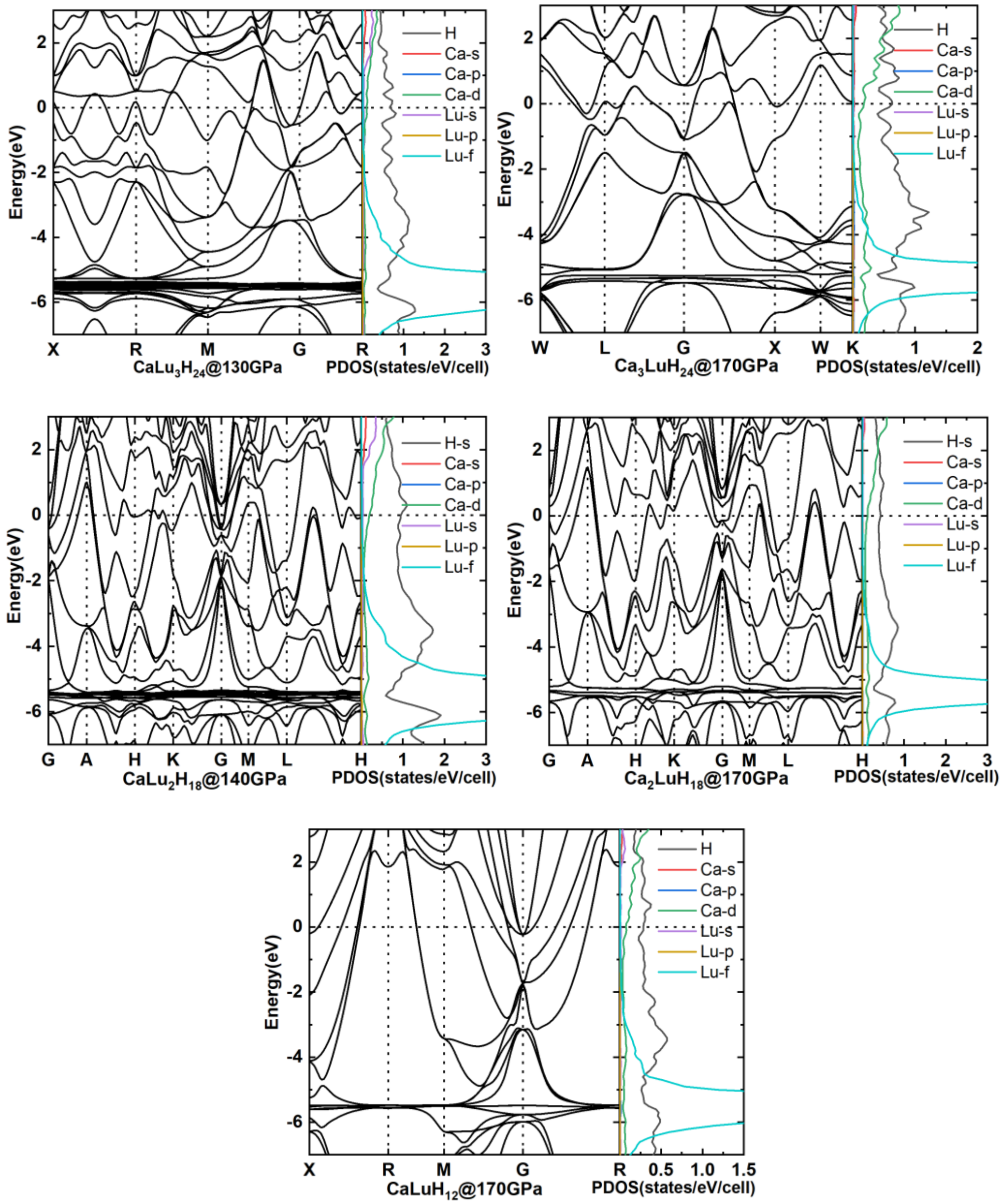

Fig. S16 Electronic band structures and projected density of electronic states of $Ca_3LuH_{24}$, $CaLu_3H_{24}$, $CaLu_2H_{18}$, $Ca_2LuH_{18}$ and $CaLuH_{12}$ under their minimum dynamically stable pressures, respectively.

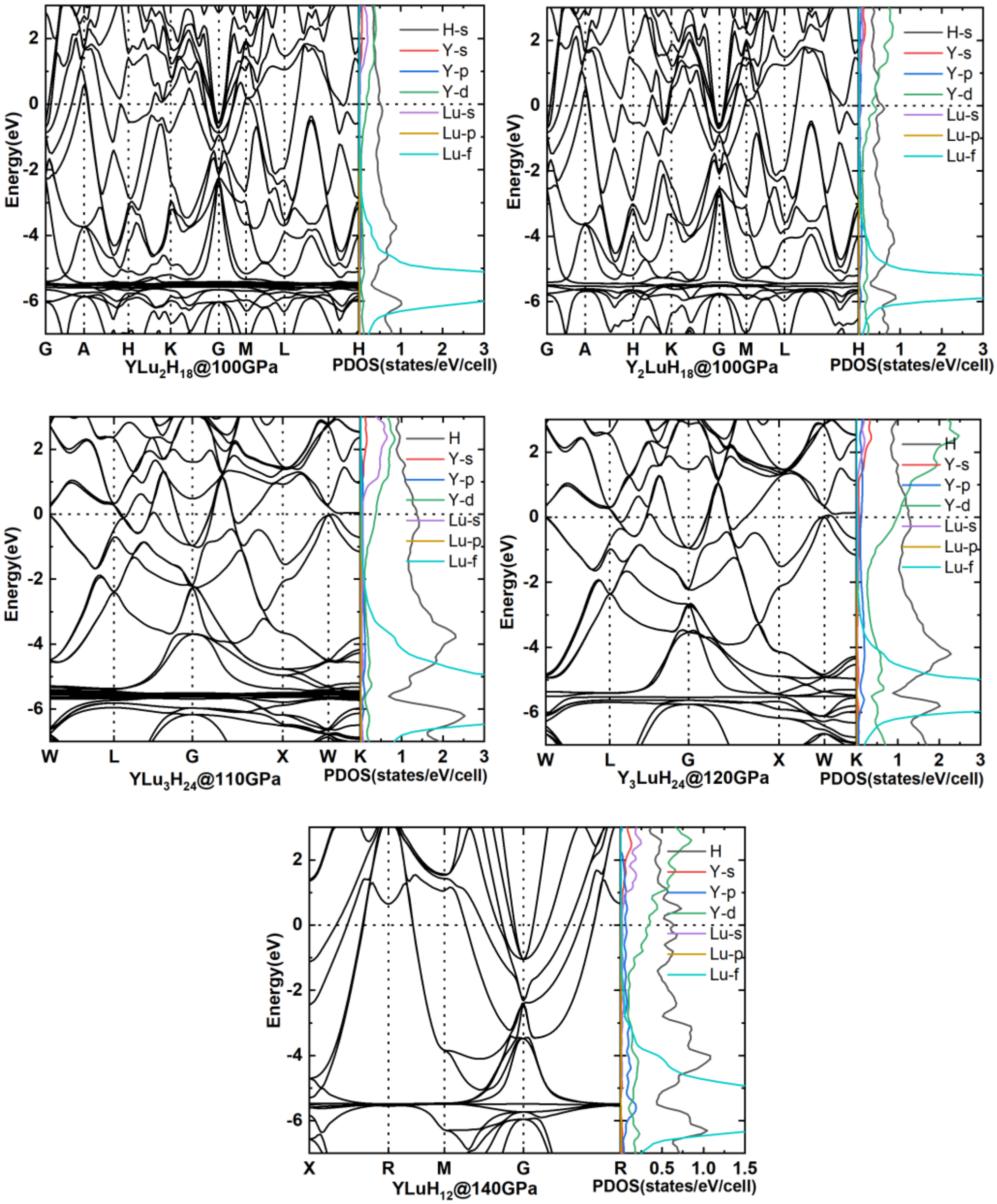

Fig. S17 Electronic band structures and projected density of electronic states of $YLu_2H_{18}$, $Y_2LuH_{18}$, $YLu_3H_{24}$, $Y_3LuH_{24}$ and $YLuH_{12}$ under their minimum dynamically stable pressures, respectively.

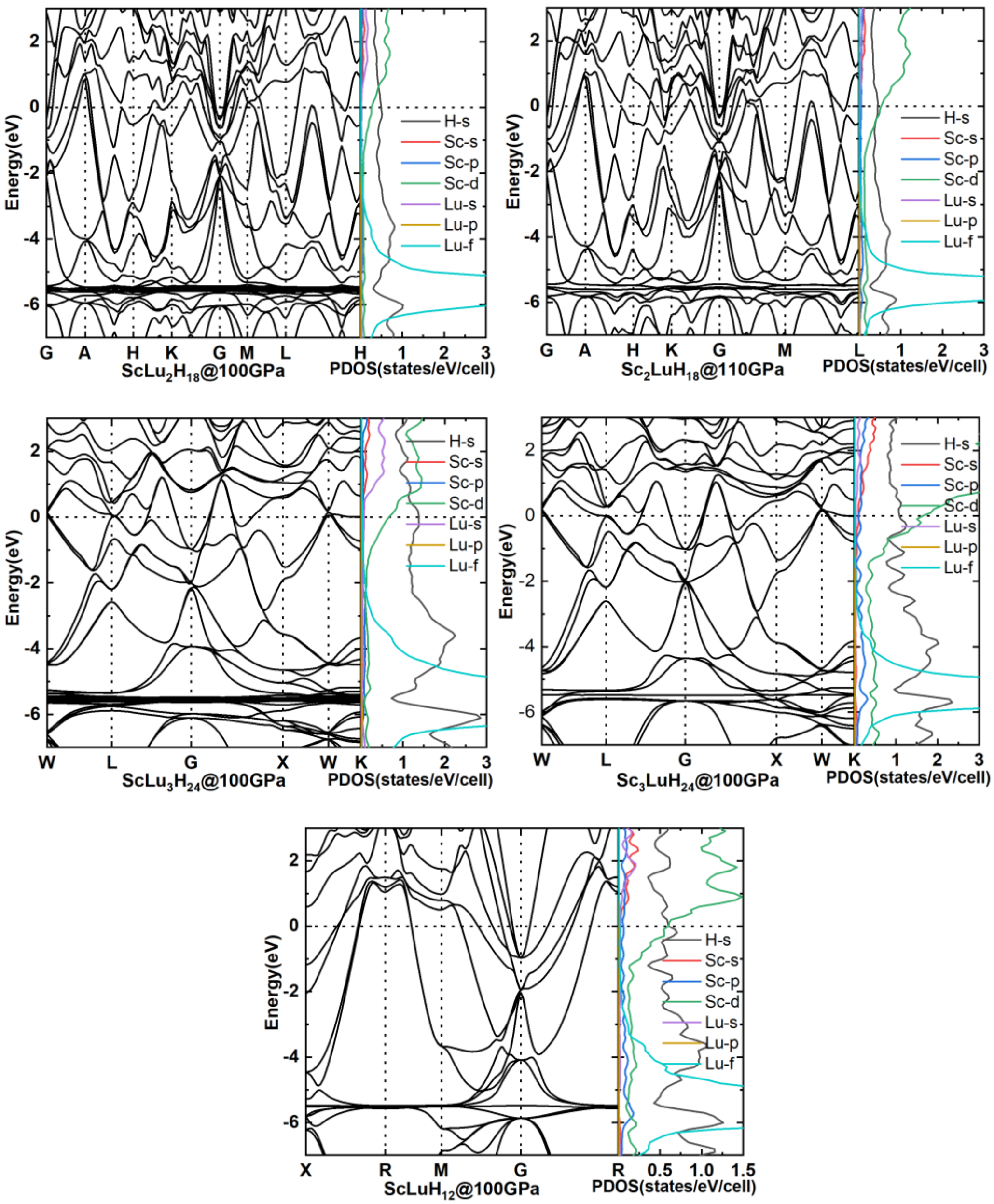

Fig. S18 Electronic band structures and projected density of electronic states of $ScLu_2H_{18}$, $Sc_2LuH_{18}$, $ScLu_3H_{24}$, $Sc_3LuH_{24}$ and $ScLuH_{12}$ under their minimum dynamically stable pressures, respectively.

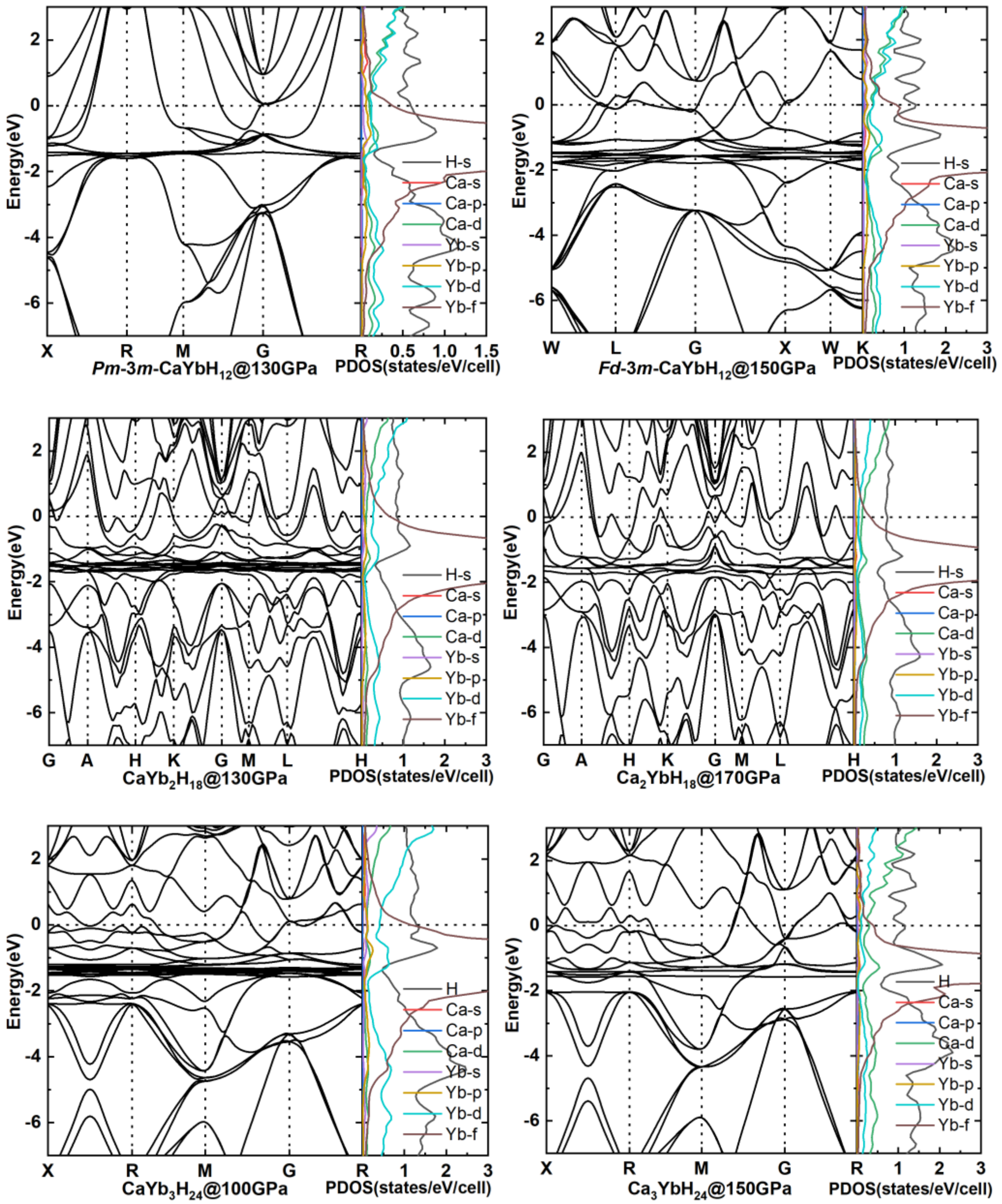

Fig. S19 Electronic band structures and projected density of electronic states of *Pm*-3*m*-$CaYbH_{12}$, *Fd*-3*m*-$CaYbH_{12}$, $CaYb_2H_{18}$, $Ca_2YbH_{18}$, $CaYb_3H_{24}$ and $Ca_3YbH_{24}$ under their minimum dynamically stable pressures, respectively.

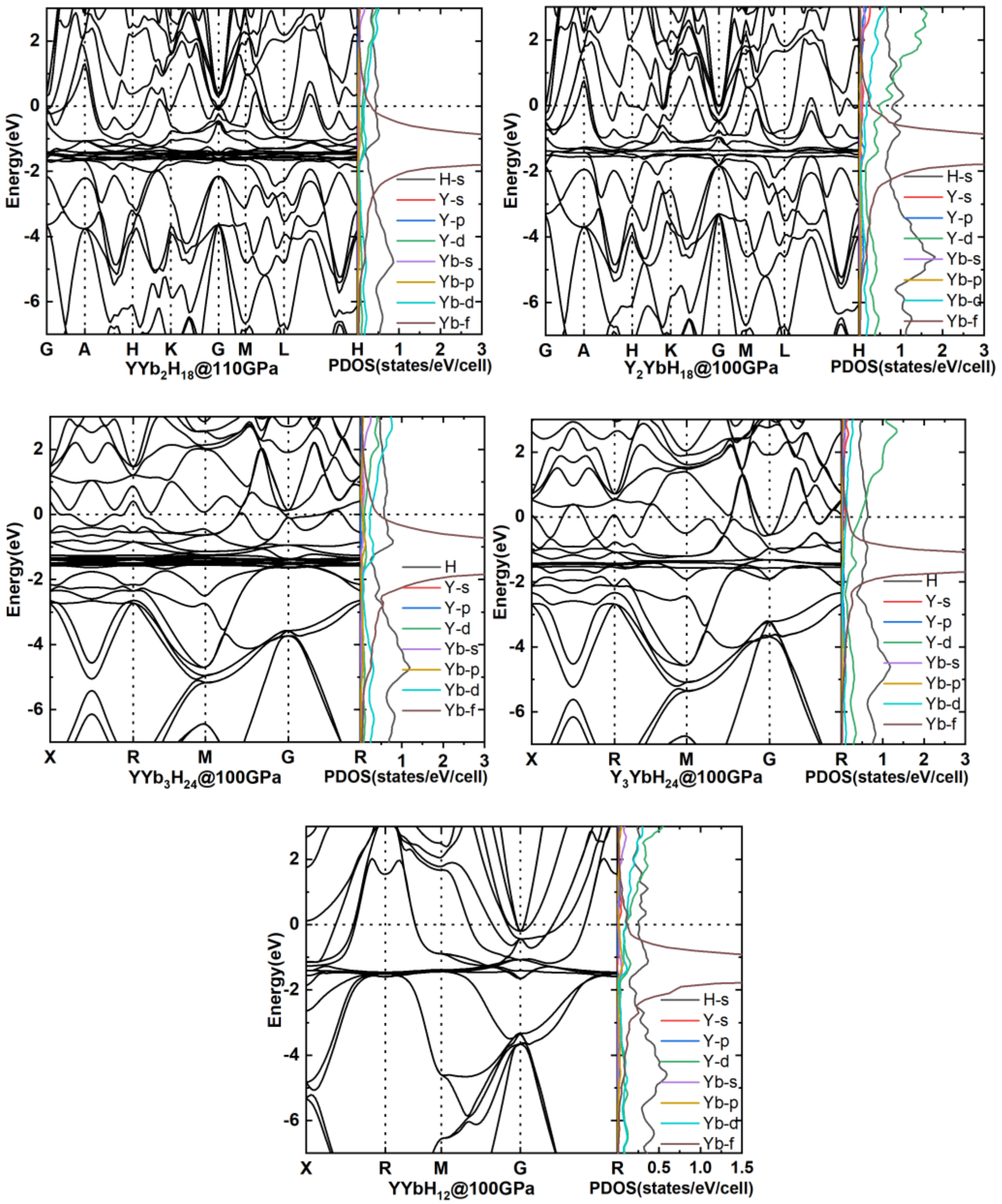

Fig. S20 Electronic band structures and projected density of electronic states of $YYb_2H_{18}$, $Y_2YbH_{18}$, $YYb_3H_{24}$, $Y_3YbH_{24}$ and $YYbH_{12}$ under their minimum dynamically stable pressures, respectively.

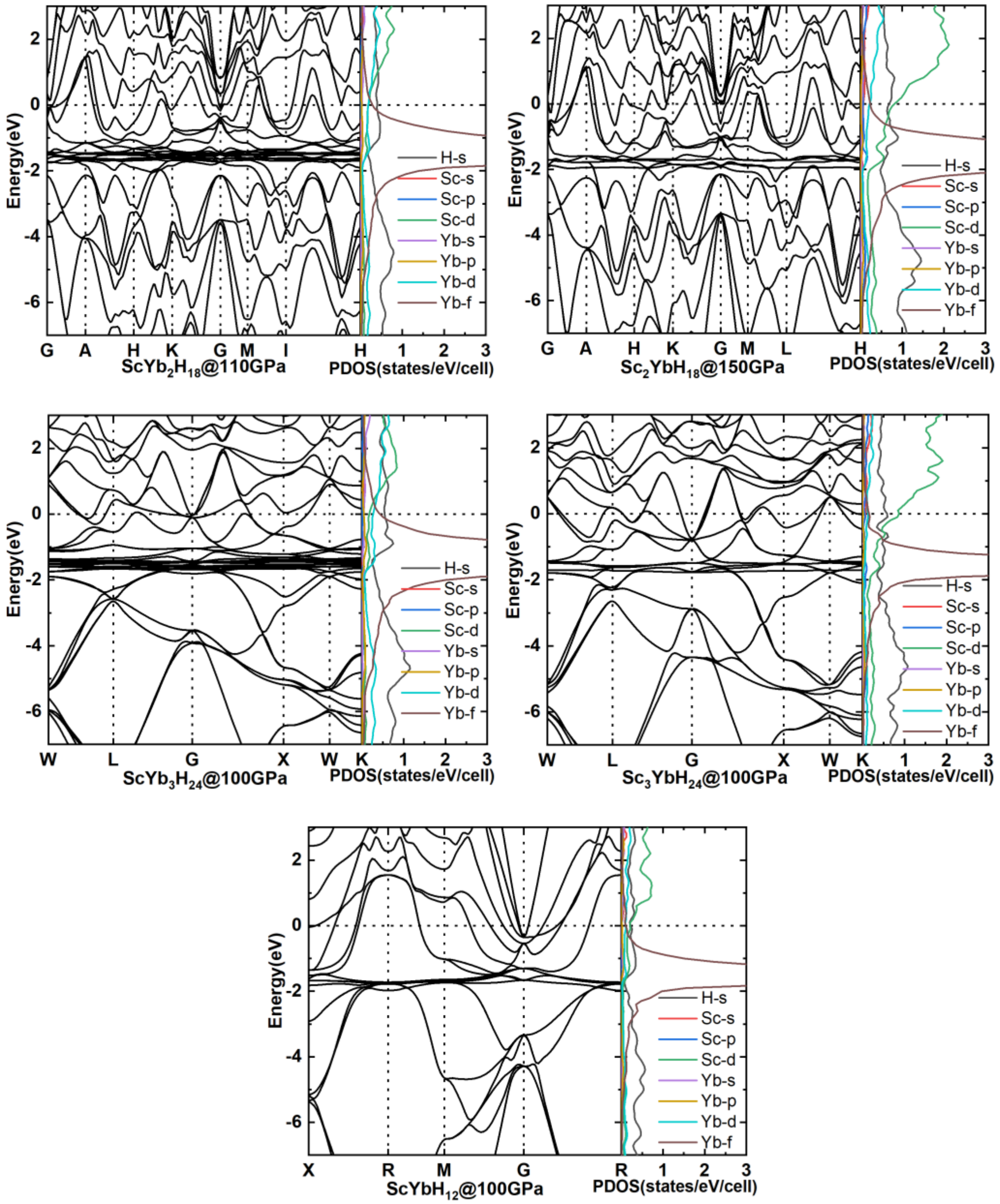

Fig. S21 Electronic band structures and projected density of electronic states of $ScYb_2H_{18}$, $Sc_2YbH_{18}$, $ScYb_3H_{24}$, $Sc_3YbH_{24}$ and $ScYbH_{12}$ under their minimum dynamically stable pressures, respectively.

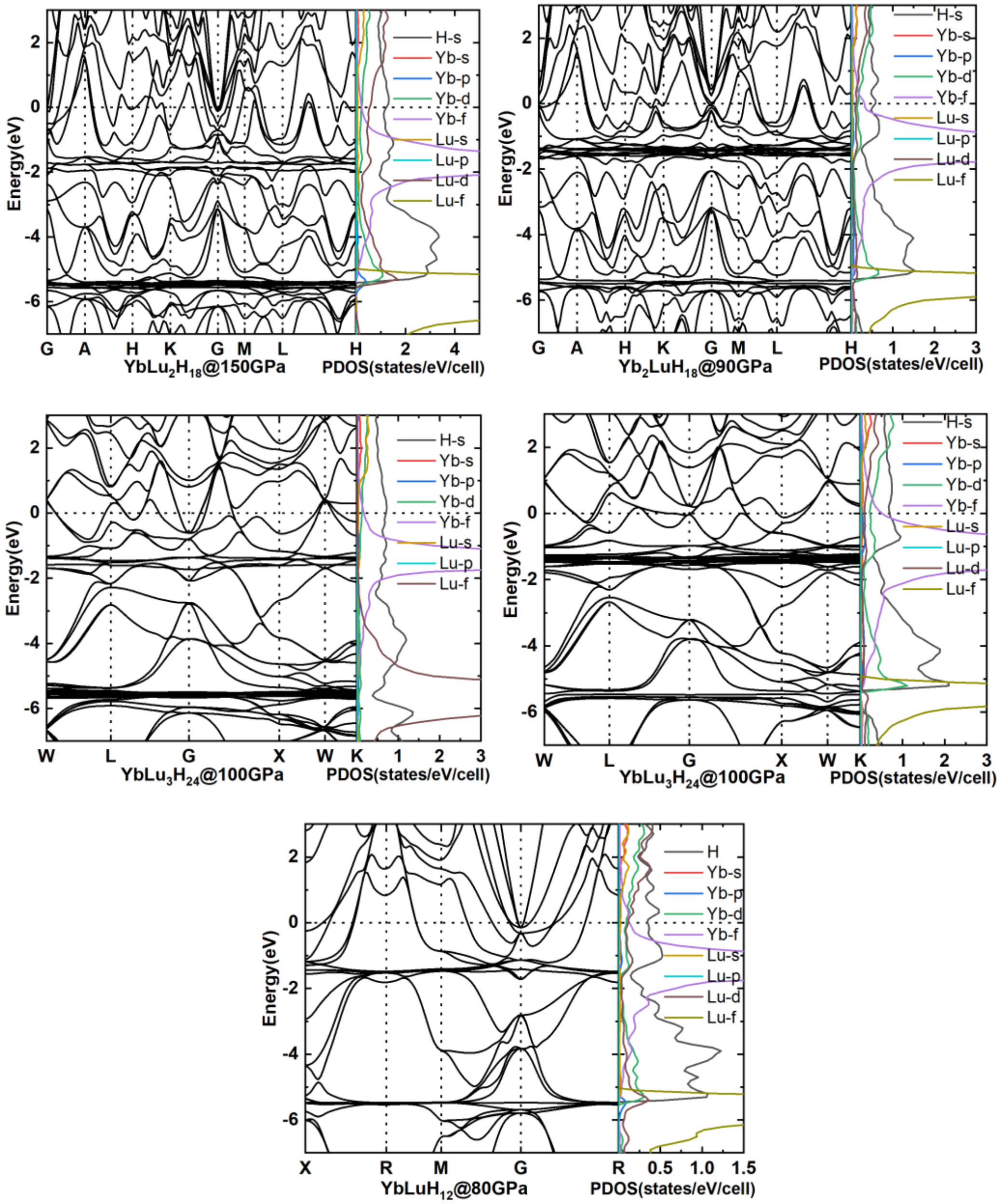

Fig. S22 Electronic band structures and projected density of electronic states of $YbLu_2H_{18}$, $Yb_2LuH_{18}$, $YbLu_3H_{24}$, $Yb_3LuH_{24}$ and $YbLuH_{12}$ under their minimum dynamically stable pressures, respectively.

## Equations for calculating $T_c$ and related parameters

### (1) The Allen−Dynes-modified McMillan equation

$T_c$ can be estimated by the McMillan equation[6]:

$$T_c = \frac{\omega_{log}}{1.2} exp\left[-\frac{1.04(1+\lambda)}{\lambda-\mu^*(1+0.62\lambda)}\right] \quad (1)$$

where λ and $\omega_{log}$ are the electron−phonon coupling constant and the logarithmic-averaged phonon frequency, respectively, and $\mu^*$ is the Coulomb pseudopotential, for which we use the widely accepted range of 0.1-0.13. λ and $\omega_{log}$ are given by

$$\lambda = 2\int_0^\infty \frac{\alpha^2F(\omega)}{\omega}d\omega \quad (2)$$

and

$$\omega_{log} = exp\left(\frac{2}{\lambda}\int_0^\infty \frac{d\omega}{\omega}\alpha^2F(\omega)\, ln\, \omega\right) \quad (3)$$

The parameter ω denotes the phonon frequency, and $\alpha^2F(\omega)$ is the Eliashberg spectral function

$$\alpha^2F(\omega) = \frac{1}{2\pi N(\varepsilon_F)}\sum_{\boldsymbol{q}v}\frac{\gamma_{qv}}{\omega_{\boldsymbol{q}v}}\delta(\omega-\omega_{\boldsymbol{q}v}) \quad (4)$$

The line width $\gamma_{q,\upsilon}$ is written as

$$\gamma_{qv} = \pi\omega_{qv}\sum_{mn}\sum_k |g_{mn}^v(\boldsymbol{k},\boldsymbol{q})|^2\,\delta(\varepsilon_{m,\boldsymbol{k}+\boldsymbol{q}}-\varepsilon_F)\times\delta(\varepsilon_{n,\boldsymbol{k}}-\varepsilon_F) \quad (5)$$

where $\varepsilon_{n,\mathbf{k}}$ is the energy of the bare electronic Bloch state, $\varepsilon_F$ is the Fermi energy, and $g_{mn}^v(\boldsymbol{k},\boldsymbol{q})$ is the electron−phonon matrix element.

When the value of $\lambda$ larger than 1.3, strong-coupling corrections begin to appear. Therefore, P. B. Allen and R. C. Dynes use two separate correction factors ($f_1$ and $f_2$) to describe these two effects. Then we can further obtain the Allen−Dynes-modified McMillan equation[7]:

$$T_c = \frac{f_1f_2\omega_{log}}{1.2} exp\left[-\frac{1.04(1+\lambda)}{\lambda-\mu^*(1+0.62\lambda)}\right] \quad (6)$$

$f_1$ and $f_2$ are given by

$$f_1 = \sqrt[3]{\left[1+\left(\frac{\lambda}{2.46(1+3.8\mu^*)}\right)^{\frac{3}{2}}\right]} \quad (7)$$

and

$$f_2 = 1+\frac{\left(\frac{\omega_2}{\omega_{log}}-1\right)\lambda^2}{\lambda^2+\left[1.82(1+6.3\mu^*)\frac{\bar{\omega}_2}{\omega_{log}}\right]} \quad (8)$$

average frequencies $\bar{\omega}_2$ is given by

$$\bar{\omega}_2 = \sqrt{\frac{2}{\lambda}\int_0^\infty \frac{d\omega}{\omega}\alpha^2F(\omega)\omega d\omega} \quad (9)$$

### (2) Self-consistent solution of the Eliashberg equation

For strong-coupling system, it can be better described with Eliashberg equation[8]:

$$Z(i\omega_n)\Delta(i\omega_n) = \frac{\pi T}{N_F}\sum_{n'}\frac{\Delta(i\omega_n')}{\sqrt{\omega_n'^2+\Delta^2(i\omega_n')}}\times[\lambda(\omega_n-\omega_{n'})-N_F\mu^*]\delta(\epsilon) \quad (10)$$

$$Z(i\omega_n) = 1 + \frac{\pi T}{N_{\mathrm{F}}\omega_n}\sum_{n'} \frac{\omega'_n}{\sqrt{\omega'^2_n + \Delta^2(i\omega'_n)}}\lambda(\omega_n - \omega_{n'})\delta(\epsilon) \tag{11}$$

where functions $Z(i\omega_n)$ and $\Delta(i\omega_n)$ are the renormalization function and pairing order parameter, respectively. $N_{\mathrm{F}}$ is the density of electronic states at the Fermi level, and $\delta(\epsilon)$ is the Dirac delta function. $i\omega_n = i(2n+1)\pi T_c$ are the fermion Matsubara frequencies. $\mu^*$ is the Coulomb pseudopotential, for which we use the widely accepted range of 0.1 - 0.13. $\lambda(\omega_n - \omega_{n'})$ contains the electron-phonon coupling matrix, phonon propagator, and the phonon density of states, and is given by:

$$\lambda(\omega_n - \omega_{n'}) = \int_0^\infty d\omega \frac{2\omega}{(\omega_n - \omega'_n)^2 + \omega^2}\alpha^2 F(\omega) \tag{12}$$

The equations for the $Z(i\omega_n)$ and $\Delta(i\omega_n)$ form a coupled nonlinear system and are solved self-consistently. We evaluated renormalization function and the order parameter for each Matsubara frequency along the imaginary energy axis. After calculating $Z(i\omega_n)$ and $\Delta(i\omega_n)$, an analytic continuation is performed to the real axis using Pade' functions.

The specific process is as follows:

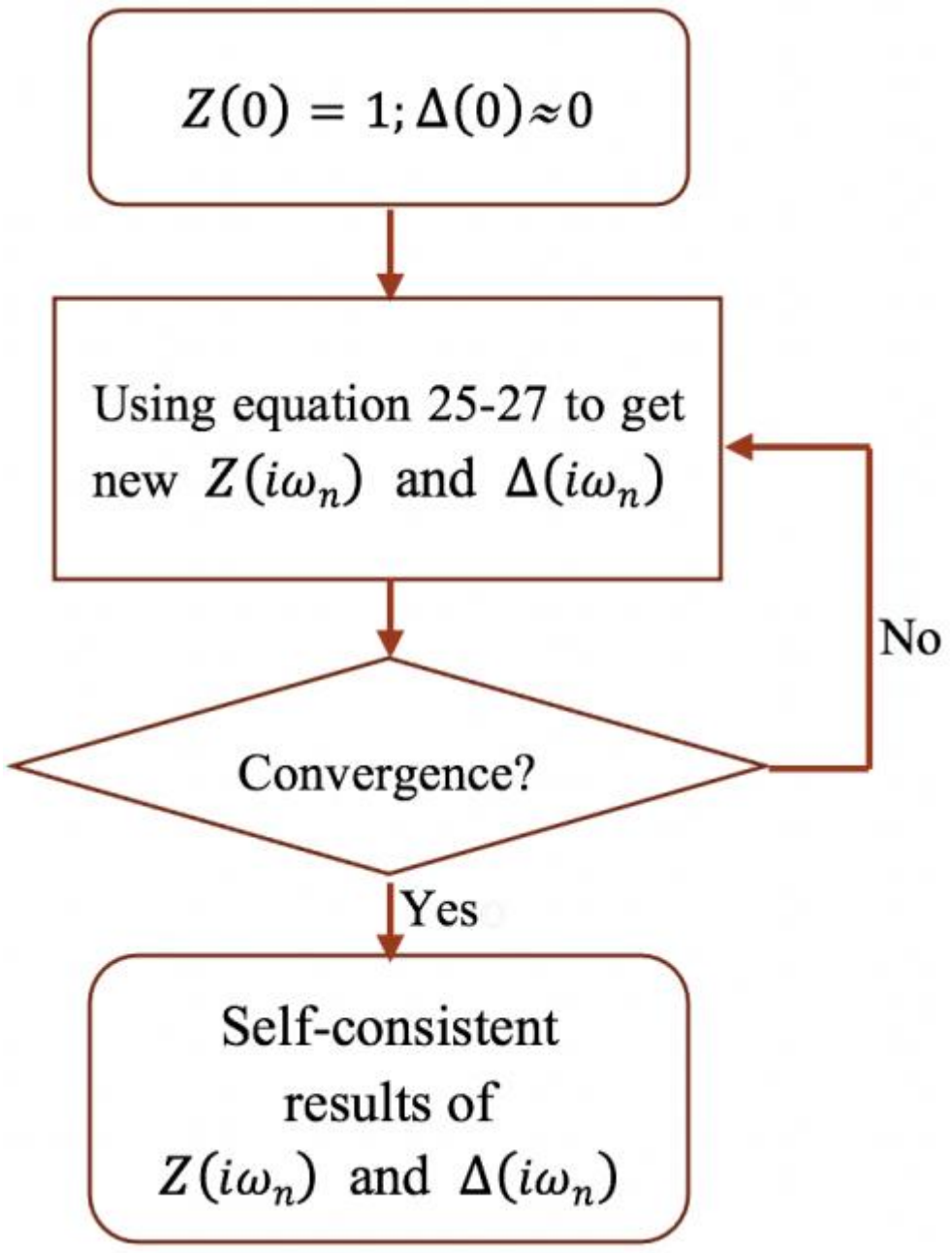

# TABLES

Table S1. The calculated electron-phonon coupling parameter λ, logarithmic average phonon frequency $\omega_{log}$, superconducting critical temperature $f_1f_2T_c$ using Allen-Dynes modified McMillan equation and $T_c^{scE}$ using the self-consistent solution of the Eliashberg equation for dynamically stable superconductors found in this work. The pressure presented here is where they start to become dynamically stable. The Coulomb pseudopotential $\mu^*$ were chose 0.10 and 0.13.

| Compound | Phase | Pressure (GPa) | λ | $\omega_{log}$ (K) | $f_1f_2T_c$ (K) | $T_c^{scE}$ (K) |
|---|---|---|---|---|---|---|
| $Ca_3LuH_{24}$ | *Fm*-3*m* | 170 | 2.64 | 816 | 169-184 | 206-221 |
| $Ca_2LuH_{18}$ | *P*-3*m*1 | 170 | 2.19 | 1087 | 188-205 | 216-232 |
| $CaLuH_{12}$ | *Pm*-3*m* | 170 | 3.94 | 648 | 197-216 | 267-282 |
| $CaLu_2H_{18}$ | *P*-3*m*1 | 140 | 4.13 | 635 | 202-222 | 284-299 |
| $CaLu_3H_{24}$ | *Fm*-3*m* | 130 | 3.60 | 840 | 198-216 | 240-255 |
| $Ca_3YbH_{24}$ | *Fm*-3*m* | 150 | 3.09 | 371 | 94-104 | 183-196 |
| $Ca_2YbH_{18}$ | *P*-3*m*1 | 170 | 2.02 | 707 | 116-128 | 158-171 |
| $CaYbH_{12}$ | *Pm*-3*m* | 130 | 3.30 | 551 | 131-144 | 165-176 |
| $CaYbH_{12}$ | *Fd*-3*m* | 150 | 1.38 | 1140 | 118-132 | 140-153 |
| $CaYb_2H_{18}$ | *P*-3*m*1 | 130 | 1.65 | 835 | 107-119 | 125-138 |
| $CaYb_3H_{24}$ | *Fm*-3*m* | 100 | 1.91 | 782 | 118-129 | 141-151 |
| $Y_3LuH_{24}$ | *Fm*-3*m* | 120 | 3.75 | 696 | 199-218 | 269-283 |
| $Y_2LuH_{18}$ | *P*-3*m*1 | 100 | 3.21 | 747 | 184-201 | 224-240 |
| $YLuH_{12}$ | *Pm*-3*m* | 140 | 3.17 | 771 | 190-208 | 256-275 |
| $YLu_2H_{18}$ | *P*-3*m*1 | 100 | 3.24 | 723 | 180-197 | 226-242 |
| $YLu_3H_{24}$ | *Fm*-3*m* | 110 | 4.78 | 495 | 183-202 | 274-288 |
| $Y_3YbH_{24}$ | *Fm*-3*m* | 100 | 4.78 | 331 | 124-138 | 209-222 |
| $Y_2YbH_{18}$ | *P*-3*m*1 | 100 | 3.24 | 595 | 151-165 | 205-221 |
| $YYbH_{12}$ | *Pm*-3*m* | 100 | 2.22 | 761 | 134-147 | 166-178 |
| $YYb_2H_{18}$ | *P*-3*m*1 | 110 | 1.76 | 946 | 131-144 | 160-172 |
| $YYb_3H_{24}$ | *Fm*-3*m* | 100 | 1.55 | 944 | 113-125 | 136-148 |
| $Sc_3LuH_{24}$ | *Fm*-3*m* | 100 | 3.58 | 484 | 139-153 | 236-250 |
| $Sc_2LuH_{18}$ | *P*-3*m*1 | 110 | 3.33 | 535 | 142-156 | 222-239 |
| $ScLuH_{12}$ | *Pm*-3*m* | 100 | 4.43 | 517 | 175-193 | 253-266 |
| $ScLu_2H_{18}$ | *P*-3*m*1 | 100 | 3.66 | 489 | 143-157 | 233-247 |
| $ScLu_3H_{24}$ | *Fm*-3*m* | 100 | 2.01 | 501 | 173-191 | 258-271 |
| $Sc_3YbH_{24}$ | *Fm*-3*m* | 100 | 2.60 | 603 | 127-139 | 190-203 |
| $Sc_2YbH_{18}$ | *P*-3*m*1 | 150 | 1.94 | 921 | 142-156 | 182-196 |
| $ScYbH_{12}$ | *Pm*-3*m* | 100 | 2.10 | 778 | 130-143 | 177-191 |
| $ScYb_2H_{18}$ | *P*-3*m*1 | 110 | 1.68 | 886 | 117-129 | 147-160 |
| $ScYb_3H_{24}$ | *Fm*-3*m* | 100 | 1.65 | 846 | 110-121 | 138-150 |
| $Yb_3LuH_{24}$ | *Fm*-3*m* | 80 | 1.97 | 787 | 123-135 | 151-162 |
| $Yb_2LuH_{18}$ | *P*-3*m*1 | 90 | 2.14 | 789 | 133-146 | 170-182 |
| $YbLuH_{12}$ | *Pm*-3*m* | 80 | 2.82 | 636 | 140-153 | 186-198 |
| $YbLu_2H_{18}$ | *P*-3*m*1 | 150 | 1.77 | 1073 | 150-166 | 197-212 |
| $YbLu_3H_{24}$ | *Fm*-3*m* | 100 | 2.65 | 833 | 171-187 | 209-222 |

Table S2. The calculated electron-phonon coupling parameter λ, logarithmic average phonon frequency $\omega_{log}$, superconducting critical temperature $f_1f_2T_c$ using Allen-Dynes modified McMillan equation and $T_c^{scE}$ using the self-consistent solution of the Eliashberg equation for $CaH_6$, $YH_6$ and $ScH_6$. The Coulomb pseudopotential $\mu^*$ were chose 0.10 and 0.13.

| Compound | Phase | Pressure (GPa) | λ | $\omega_{log}$ (K) | $f_1f_2T_c$ (K) | $T_c^{scE}$ (K) |
|---|---|---|---|---|---|---|
| $CaH_6$ | *Im*-3*m* | 150 | 2.46 | 1050 | 195-212 | 220-233 |
| $CaH_6$ | *Im*-3*m* | 172 | 2.06 | 1167 | 186-203 | 209-223 |
| $CaH_6$ | *Im*-3*m* | 172 | exp[9] | | 215 | |
| $YH_6$ | *Im*-3*m* | 120 | 2.77 | 866 | 186-203 | 232-247 |
| $YH_6$ | *Im*-3*m* | 166 | 1.98 | 1154 | 180-198 | 218-234 |
| $YH_6$ | *Im*-3*m* | 166 | exp[10] | | 224 | |
| $ScH_6$ | *Im*-3*m* | 130 | 2.26 | 752 | 137-150 | 189-203 |

Table S3 The lattice parameters and atomic positions of $YLuH_{12}$, $YLu_2H_{18}$, $Y_2LuH_{18}$, $YLu_3H_{24}$ and $Y_3LuH_{24}$ under different pressures.

| **Structure** | **Parameters** (Å, deg) | **Atom** | **x** | **y** | **z** |
|---|---|---|---|---|---|
| *Pm*-3*m* | a=b=c= 3.6094 | H | 0.25322 | 0.00000 | 0.50000 |
| $YLuH_{12}$ | α=β=γ= 90 | Y | 0.50000 | 0.50000 | 0.50000 |
| (140 GPa) | | Lu | 0.00000 | 0.00000 | 0.00000 |
| *Fd*-3*m* | a=b=c= 4.9396 | H | 1.37371 | 0.12500 | -0.37500 |
| $YLuH_{12}$ | α=β=γ= 60 | Y | 1.75000 | -0.25000 | -0.25000 |
| (200 GPa) | | Lu | 1.00000 | 0.00000 | 0.00000 |
| *P*-3*m*1 | a=b= 5.2419 | H | 0.08731 | 0.66858 | 0.84076 |
| $YLu_2H_{18}$ | c= 3.2088 | H | 0.25752 | 0.25752 | 0.50000 |
| (100 GPa) | α=β= 90 | Y | 0.00000 | 0.00000 | 0.00000 |
| | γ= 120 | Lu | 0.66667 | 0.33333 | 0.65774 |
| *P*-3*m*1 | a=b= 5.2633 | H | 0.58588 | 0.66623 | 0.66995 |
| $Y_2LuH_{18}$ | c= 3.2244 | H | 0.24566 | 0.24566 | 0.00000 |
| (100 GPa) | α=β= 90 | Y | 0.66667 | 0.33333 | 0.83002 |
| | γ= 120 | Lu | 0.00000 | 0.00000 | 0.50000 |
| *Fm*-3*m* | a=b=c= 5.1959 | H | -0.87486 | 1.87486 | -0.37363 |
| $YLu_3H_{24}$ | α=β=γ= 60 | Y | -0.50000 | 1.50000 | -0.50000 |
| (110 GPa) | | Lu | -0.25000 | 0.75000 | -0.25000 |
| | | Lu | 0.00000 | 0.00000 | 0.00000 |
| *Fm*-3*m* | a=b=c= 5.1894 | H | 1.87502 | -0.62340 | -0.87502 |
| $Y_3LuH_{24}$ | α=β=γ= 60 | Y | 0.00000 | 0.00000 | 0.00000 |
| (120 GPa) | | Y | 2.25000 | -0.75000 | -0.75000 |
| | | Lu | 1.50000 | -0.50000 | -0.50000 |